\documentclass[%
 reprint,
 amsmath,amssymb,
 aps,
]{revtex4-2}

\usepackage{graphicx}
\usepackage{dcolumn}
\usepackage{bm}

\usepackage{bigstrut}
\usepackage{color}
\usepackage[normalem]{ulem}

\usepackage{color}

\begin{document}

\preprint{APS/123-QED}

\title{Opinion dynamics on tie-decay networks}

\author{Kashin Sugishita}
\thanks{Present address: Department of Transdisciplinary Science and Engineering, Tokyo Institute of Technology, 152-8550 Tokyo, Japan}
\affiliation{%
Department of Mathematics, State University of New York at Buffalo, Buffalo, New York 14260-2900, USA}

\author{Mason A. Porter}
\affiliation{
Department of Mathematics, University of California, Los Angeles, California 90095, USA
}%

\author{Mariano Beguerisse-D\'{i}az}
\thanks{Present address: Spotify Ltd, 4 Savoy Place, London WC2N 6AT, UK}
\affiliation{%
 Mathematical Institute, University of Oxford, Oxford  OX2 6GG, UK
}

\author{Naoki Masuda}
 \email{naokimas@buffalo.edu}
\affiliation{%
Department of Mathematics, State University of New York at Buffalo, Buffalo, New York 14260-2900, USA}
\affiliation{
Computational and Data-Enabled Science and Engineering Program, State University of New York at Buffalo, Buffalo, New York 14260-5030, USA}
\affiliation{Faculty of Science and Engineering, Waseda University, 169-8555 Tokyo, Japan}

\date{\today}

\begin{abstract}

In social networks, interaction patterns typically change over time. We study opinion dynamics on tie-decay networks in which tie strength increases instantaneously when there is an interaction and decays exponentially between interactions. Specifically, we formulate continuous-time Laplacian dynamics and a discrete-time DeGroot model of opinion dynamics on these tie-decay networks, and we carry out numerical computations for the continuous-time Laplacian dynamics. We examine the speed of convergence by studying the spectral gaps of combinatorial Laplacian matrices of tie-decay networks. First, we compare the spectral gaps of the Laplacian matrices of tie-decay networks that we construct from empirical data with the spectral gaps for corresponding randomized and aggregate networks. We find that the spectral gaps for the empirical networks tend to be smaller than those for the randomized and aggregate networks. Second, we study the spectral gap as a function of the tie-decay rate and time. Intuitively, we expect small tie-decay rates to lead to fast convergence because the influence of each interaction between two nodes lasts longer for smaller decay rates. Moreover, as time progresses and more interactions occur, we expect eventual convergence. However, we demonstrate that the spectral gap need not decrease monotonically with respect to the decay rate or increase monotonically with respect to time. Our results highlight the importance of the interplay between the times that edges strengthen and decay in temporal networks.
\end{abstract}
\maketitle




\section{Introduction}\label{sec:Introduction}

One can represent the structure of many natural, societal, and engineered systems as networks \citep[][]{newman2018networks}. The simplest type of network is a graph, which consists of nodes (such as individual humans) and edges, each of which connects two nodes to each other. The qualitative dynamics of many types of dynamical processes that consist of interacting elements (i.e., nodes) --- including the spread of infectious diseases, opinion formation, synchronization, and cascading failures in power grids --- depend considerably on network structure \cite{barrat2008dynamical,boccaletti2006complex,porter2016dynamical}.

In the present paper, we consider models of opinion dynamics. People in a social network have different opinions, and the opinions of people can change as they influence each other \cite{noor2020,castellano2009statistical}. Examples include choosing between candidates in an election, debating topics in online forums, and collective decision-making in animal flocks. Models of opinion dynamics aim to explore the emergence of consensus, persistent disagreement, transitions between consensus and disagreement, the influence of the heterogeneity of individuals, the effects of media on opinion dynamics, and more \cite{baronchelli2018emergence, barrat2008dynamical, boccaletti2006complex,porter2016dynamical,castellano2009statistical,sen2014sociophysics,sirbu2017opinion,acemouglu2013opinion,brooks2020}.
Many previous studies have examined opinion dynamics on time-independent networks and have explored rich phenomena, such as the probability to reach consensus and the time to reach it, the emergence of different opinion clusters, the influence of stubborn individuals on persistent disagreement, and the effects of network structure on these phenomena \citep[][]{noor2020,castellano2009statistical, sen2014sociophysics, sirbu2017opinion}. However, most social networks change in time \cite{holme2012temporal,holme2015modern,holme2019temporal}. Individuals move from one place to another during their lives, friendships form and dissipate, interactions differ in different times of a day (e.g., morning versus night) and different days of a week (e.g., weekdays versus weekends), and so on.
This yields time-varying networks, which are often called ``temporal networks'' \citep[][]{holme2015modern, holme2012temporal, holme2019temporal,masuda2016guidance}. It is important to examine dynamical processes, such as opinion dynamics \cite{fernandez2011update, masuda2013temporal, olfati2007consensus, takaguchi2013bursty}, on temporal networks.
 In such scenarios, both the structure of a network and the states of its nodes and/or edges change in time.

Although many time-dependent networked systems evolve continuously in time, it is a common practice to aggregate interactions into homogeneous and discrete time windows to facilitate the analysis of such systems \cite{holme2019temporal, masuda2016guidance}. 
If one is not careful about examining the relative time scales of dynamical processes on a network and temporal changes in network structure (or if multiple time scales or burstiness are relevant to one or more of these processes), then aggregating network dynamics into discrete time windows may lead to qualitatively incorrect conclusions. Such situations have motivated investigations of the size of aggregation time windows and the development of methods to allow them to be heterogeneous \citep[][]{caceres2011temporal, krings2012effects,liljeros2007contact,psorakis2012inferring}. 
An alternative approach that avoids the use of discrete time windows is to consider time-stamped events, such as conversation events, between nodes as the objects of interest \cite{holme2015modern,holme2012temporal, masuda2016guidance}. Because events between nodes are often bursty in social networks \cite{beguerisse2010competition, holme2012temporal, karsai2018bursty, kivela2015estimating, moinet2015burstiness}, it is critical to carefully consider both the sizes and the boundaries of time windows.

Despite the wealth of previous studies of temporal networks, we still lack robust and principled frameworks to study temporal network data that comes in the form of a list or stream of time-stamped events. A promising framework for studying temporal networks in continuous time is ``tie-decay networks'' \cite{ahmad2018tie, zuo2019models}, in which one distinguishes the concepts of interactions and ties. Interactions represent discrete contacts, and ties represent relationships between entities that change continuously in time. The strength of a tie decays in time and increases instantaneously by some amount when there is an interaction event. Importantly, tie-decay networks do not impose a hard partitioning of the set of interaction events into discrete time windows. In the context of opinion dynamics, the use of a tie-decay network entails that the effect of each interaction event on the opinions of the two nodes in the interaction lasts for some duration after the event.

In the present paper, we study opinion dynamics on tie-decay networks. We examine continuous-time Laplacian dynamics \cite{mirzaev2013laplacian} and a discrete-time DeGroot model of opinion dynamics \cite{urena2019review} on such networks, and we conduct numerical computations for the continuous-time Laplacian dynamics. We examine the convergence speed of opinion dynamics that arise from a time-varying combinatorial Laplacian matrix for tie-decay networks that we construct from empirical data. We also compare the convergence speed of opinion dynamics on the original tie-decay networks to such dynamics on randomized and aggregate networks. We find that the convergence speeds of opinion dynamics on the tie-decay networks that we construct from the empirical data tend to be slower than the convergence speeds on the associated randomized and aggregate networks. Interestingly, we also find that the convergence speed need not decrease monotonically with respect to the decay rate and that the convergence need not proceed monotonically with respect to time.

Our paper proceeds as follows. In Sec.~\ref{sec:formulate}, we formulate continuous-time Laplacian dynamics and a discrete-time DeGroot model of opinion dynamics on tie-decay networks. In Sec.~\ref{sec:random_aggregate}, we describe how we generate randomized and aggregate networks from a given temporal network. In Sec.~\ref{sec:data_sets}, we describe the six empirical data sets that we examine. In Sec.~\ref{sec:results_randomized_aggregate}, we compare the spectral gaps between these empirical tie-decay networks, randomized tie-decay networks, and aggregate networks. In Sec.~\ref{sec:non_monotonicity}, we examine the spectral gaps as a function of the decay rate and decay time. In Sec.~\ref{sec:conclusions}, we conclude and discuss the implications of our work. We discuss some additional details in three appendices.


\section{Opinion dynamics on tie-decay networks}
\label{sec:formulate}

We consider both continuous-time and discrete-time opinion dynamics on tie-decay networks. Specifically, we formulate continuous-time Laplacian dynamics and a discrete-time DeGroot model of opinion dynamics on tie-decay networks. When we track observations of the continuous-time Laplacian dynamics at the discrete times at which events occur, we obtain a DeGroot model.


\subsection{Tie-decay networks and opinion dynamics}
\label{sec:tie-decay}

We start by briefly describing tie-decay networks \cite{ahmad2018tie}. By using tie-decay networks, we distinguish between interactions (which represent discrete contacts) and ties (which represent relationships that change continuously in time). Suppose that there are $N$ nodes, and let $B(t)$ be an $N\times N$ time-dependent matrix with real, non-negative entries $b_{ij}(t)$ for all node pairs $(v_i,v_j)$.
The entry $b_{ij}(t)$ encodes the strength of the {tie from node $v_i$ to node $v_j$} at time $t$. 
The tie strength (i.e., edge weight) $b_{ij}(t)$ evolves according to the following two rules. First, in the absence of interactions, ties decay exponentially in time; specifically $\frac{{\rm d}b_{ij}}{{\rm d}t}=-\alpha b_{ij}$, where $\alpha > 0$ is the decay rate. 
{We use the same decay rate for all edges.} Second, if an event occurs on the edge from $v_i$ to $v_j$ at time $t$, then the tie strength $b_{ij}(t)$ grows instantaneously by $1$.

In the following subsections, we formulate opinion dynamics on tie-decay networks. We assume that each node has a real-valued opinion that changes continuously in time. Each node $v_i$ updates its opinion such that (1) the magnitude of the opinion change in one unit of time is the sum of the influence from all of its in-neighbors, (2) the influence of each in-neighbor $v_j$ on $v_i$ depends on the difference between $v_i$'s opinion and $v_j$'s opinion, and (3) the influence of $v_j$ on $v_i$ is proportional to the tie strength $b_{ji}(t)$.


\subsection{Laplacian dynamics}
\label{sec:laplacian_dynamics}

Let $L(t)$ denote the combinatorial Laplacian matrix, which we construct from the set of the events at time $t$. For $i\neq j$, the $(i, j)$th entry $L_{ij}(t)$ of $L(t)$ is $-1$ if there is a directed edge from node $v_i$ to node $v_j$; otherwise, $L_{ij}(t)=0$. The diagonal element $L_{ii}(t)$ is equal to the out-degree of node $v_i$ for each $i \in \{1, \ldots, N\}$. Let $\Tilde{L}({t^+})$ denote the combinatorial Laplacian matrix of the tie-decay network at time $t{^+}$. 
Mathematical objects at times {$t^-$ and $t^+$} refer, respectively, to the situations immediately before and immediately after the occurrence of events at time $t$.  
Therefore, the matrix $\Tilde{L}({t^+})$ includes the effects of the events at time $t$ and those at times $t' < t$. Given the combinatorial Laplacian matrix of a tie-decay network at time $t_{n-1}$ and the events at time $t_n$, the dynamics of the combinatorial Laplacian matrix of the tie-decay network satisfy
\begin{equation}
	\Tilde{L}_{ij}({t_n^+})=\Tilde{L}_{ij}({t_{n-1}^+}) \exp\left[{-\alpha (t_n-t_{n-1})}\right] + L_{ij}(t_n)\,.
\label{eq:laplacian_dynamics}
\end{equation}

Suppose that the initial events in a network occur at time $t_0 = 0$ and that the next events (on some edges) occur at time $t_1$. This yields
\begin{equation}
	\Tilde{L}({t^-}) = \Tilde{L}({0^+}) e^{-\alpha t}\,,
\end{equation}
where $\Tilde{L}({0^+})=L(0)$ and $0\le t < t_1$.

Let $\bm x$ denote the $N$-dimensional row vector of the opinions of the nodes in our tie-decay network. Assume that $\bm x$ obeys combinatorial Laplacian dynamics in continuous time:
\begin{eqnarray}
	\frac{{\rm d}\bm x}{{\rm d}t} = -\bm x\Tilde{L}(t)^{\top}  = -\bm x \tilde{L}({0^+})^{\top} e^{-\alpha t} \nonumber\\ (\text{with } 0\le t< t_1)\,,
\label{eq:Laplacian dyn}
\end{eqnarray}
where $^\top$ represents transposition. The solution to Eq.~\eqref{eq:Laplacian dyn} is 
\begin{eqnarray}
	\bm x(t) = \bm x(0)\exp\left[\frac{\tilde{L}({0^+})^{\top}}{\alpha}(e^{-\alpha t}-1)\right] \nonumber\\ (\text{with } 0\le t< t_1)\,.
\label{eq:x(t) Laplacian dyn}
\end{eqnarray}
However, we have a stream of events, so we also need to consider other time intervals. Therefore, we need a more general version of Eq.~\eqref{eq:x(t) Laplacian dyn} for the time intervals $t_1\le t < t_2$, $t_2 \le t < t_3$, $\ldots$\,, $t_{n-1} \le t < t_n$. We write
\begin{align}
	\bm x(t) &= \bm x(t_{n'-1})\exp\left[\frac{\tilde{L}({t_{n'-1}^+})^{\top}}{\alpha}(e^{-\alpha (t-t_{n'-1})}-1)\right] \nonumber\\ &\qquad (\text{with } t_{n'-1}\le t< t_{n'}\,, \, n' \in \{1, 2, \ldots\})\,.
\label{eq:x(t) Laplacian dyn_new}
\end{align}
Using Eq.~\eqref{eq:x(t) Laplacian dyn_new}, we obtain
\begin{equation}
	\bm x(t_n) = \bm x(0) M(t_n)\,,
\label{eq:x_M}
\end{equation}
where
\begin{align}
	M(t_n) = &\exp\left[\frac{\tilde{L}({0^+})^{\top}}{\alpha}(e^{-\alpha t_1}-1)\right] \nonumber\\&\times \exp\left[\frac{\tilde{L}({t_1^+})^{\top}}{\alpha}(e^{-\alpha (t_2-t_1)}-1)\right] \times \cdots \nonumber\\&\times
\exp\left[\frac{\tilde{L}({t_{n-1}^+})^{\top}}{\alpha}(e^{-\alpha (t_n-t_{n-1})}-1)\right]\,.
\label{eq:finite_product}
\end{align}
The matrix $M(t_n)$ has an eigenvalue of $1$ because each exponential matrix on the right-hand side of Eq.~\eqref{eq:finite_product} has an eigenvalue of $1$ with corresponding left eigenvector $(1, \ldots, 1)$. We do not need to distinguish between $t_n^+$ and $t_n^-$ for $\bm x(t_n)$ and $M(t_n)$ because these quantities are continuous in time.


One can also express Eq.~\eqref{eq:finite_product} in terms of streaming data. Given the opinions of the nodes at time $t_n$ 
and the combinatorial Laplacian matrix of a tie-decay network at time ${t_n^+}$,
 the state at time $t_{n+1}$ 
 is 
\begin{align}
	\bm x(t_{n+1}) =\bm x(t_n) 
	\times \exp\left[\frac{\tilde{L}({t_n^+})^{\top}}{\alpha}\left(e^{-\alpha (t_{n+1}-t_n)}-1\right)\right] \,.
\label{eq:finite_product_2}
\end{align}


\subsection{DeGroot model of opinion dynamics}
\label{sec:Degroot_model}

We use the discrete-time DeGroot model \citep{FB-LNS, degroot1974reaching} as a model of opinion dynamics. We consider discrete time steps of length $\Delta t$. Let $\bm y=(y_1, \ldots, y_N)$ denote the $N$-dimensional row vector of the opinions of the nodes. Because time is discrete, we examine $\bm y(n \Delta t)$ with $n=0, 1, 2, \ldots$.

Let $A(n\Delta t)$ denote the adjacency matrix that we construct from the set of events at time $n \Delta t$, and let $\Tilde{A}(n\Delta t)$ denote the adjacency matrix of the tie-decay network at time $n \Delta t$. The matrix $\Tilde{A}(n\Delta t)$ includes the contributions of the events at time $n\Delta t$ and the effect of the {exponential} decay of the ties in $A(n' \Delta t)$ for $n' < n$. Given the adjacency matrix at $t=(n-1)\Delta t$ and the events at time $t=n \Delta t$, the dynamics of the adjacency-matrix elements of the tie-decay network are
\begin{equation}
	\Tilde{A}_{ij}(n\Delta t)=\Tilde{A}_{ij}((n-1)\Delta t) e^{-\alpha \Delta t} + A_{ij}(n\Delta t)\,,
\label{eq:DeGroot_1}
\end{equation}
with the convention that $\Tilde{A}_{ij}(-\Delta t) = 0$. 
The update rule in the DeGroot model is 
\begin{equation}
	y_{i}((n+1) \Delta t)=\sum_{j=1}^{N}y_{j}(n \Delta t)\Tilde{B}_{ji}((n+1)\Delta t) \,,
\label{eq:DeGroot_2}
\end{equation}
where
\begin{equation}
	\Tilde{B}_{ij}(n\Delta t)=\frac{\Tilde{A}_{ij}(n\Delta t)}{\sum_{i=1}^{N} \Tilde{A}_{ij}(n\Delta t) }\,.
\label{eq:DeGroot_3}
\end{equation}

Equation \eqref{eq:DeGroot_2} implies that 
\begin{equation}
	\bm y(n \Delta t)=\bm y_{\text{init}}\tilde{B}(0)\tilde{B}(1)\cdots  \tilde{B}(n\Delta t)\,,
\label{eq:DeGroot_5}
\end{equation}
where $\bm y_{\text{init}}$ is the initial condition before streaming network data arrives at time $t=0$.

The matrix $\tilde{B}(0)\tilde{B}(1) \cdots \tilde{B}(n\Delta t)$ has an eigenvalue of $1$ because each matrix $\tilde{B}(n^{\prime} \Delta t)$ (with $n^{\prime}\in \{0, 1, \ldots, n\}$) has an eigenvalue of $1$ with corresponding left eigenvector $(1, \ldots, 1)$.

By writing the  continuous-time Laplacian dynamics in terms of the composite map~\eqref{eq:finite_product}, we see that it is in fact a DeGroot model. The correspondence between Eqs.~\eqref{eq:finite_product} and \eqref{eq:DeGroot_5} is given by
\begin{equation}
	\tilde{B}(n) = \exp \left[\frac{\tilde{L}({t_{n-1}^+})^{\top}}{\alpha}(e^{-\alpha (t_n-t_{n-1})}-1)\right]\,.
\label{eq:laplacian_degroot}
\end{equation}
The column sum of the matrix on the right-hand side of Eq.~\eqref{eq:laplacian_degroot} is $1$ for each column. This is consistent with the normalization $\sum_{i=1}^N \tilde{B}_{ij} = 1$ (with $j \in \{1, \ldots, N\}$), which follows from Eq.~\eqref{eq:DeGroot_3}.

\begin{table*}[t]
\caption{\label{tab:four_randomization}Comparison of the randomization methods. (The symbol `P' indicates that a property is preserved, and the symbol `D' indicates that a property is destroyed.)
} 
\begin{ruledtabular}
\begin{tabular}{lcccc}
& \begin{tabular}[c]{@{}c@{}} Temporal \\
                      correlations\end{tabular} & \begin{tabular}[c]{@{}c@{}}Distribution of inter-event\\ times of each edge\end{tabular} & \begin{tabular}[c]{@{}c@{}} Number of events \\at each time step\end{tabular} & \begin{tabular}[c]{@{}c@{}}Structure of \\aggregate network\end{tabular}  \\ \hline
Interval shuffling    & D                                                                  & P                                                                    & D                                                                               & P                                       \\
Shuffled time stamps  & D                                                                  &D                                                                    & P                                                                               & P    \\
Random times          & D                                                                  & D                                                                   & D                                                                               & P  \\
Random edge shuffling & D                                                                 & D                                                                   & P                                                                              & D                                                         
\\
\end{tabular}
\end{ruledtabular}
\end{table*}

\section{Generation of randomized and aggregate networks}
\label{sec:random_aggregate}

{We compare the convergence speeds of the opinion dynamics on empirical temporal networks with those in corresponding randomized temporal networks. Randomized temporal networks are useful for investigating the effect of the properties of empirical temporal networks. If empirical and associated randomized temporal networks yield different convergence speeds,
we gain insight into the properties of empirical temporal networks that are likely responsible for such differences.
We also compare the convergence speeds of opinion dynamics on the empirical temporal networks with those in the corresponding aggregate networks. 
If an empirical temporal network and its associated aggregate network yield different convergence speeds, it is likely that the aggregation loses essential information about the opinion dynamics.}

We now discuss how we generate randomized and aggregate networks. In this section and the following sections, we consider only undirected networks.


\subsection{Randomized networks}

We investigate the convergence speeds of continuous-time Laplacian dynamics on several tie-decay networks and compare them to the convergence speeds for corresponding randomized networks. We consider four of the many ways to randomize temporal networks \citep{gauvin2018randomized}.
We explain each randomization method and summarize them in Table \ref{tab:four_randomization}.

\begin{table*}
\caption{\label{tab:original_networks}Summary of the original networks from the six data sets.}
\begin{ruledtabular}
\begin{tabular}{lrrrrr}
          Data set     & Nodes & Edges & \begin{tabular}[c]{@{}c@{}}Number\\ of events\end{tabular}& \begin{tabular}[c]{@{}c@{}}Mean number of\\ events per node\end{tabular} & \begin{tabular}[c]{@{}c@{}}Time\\ resolution [seconds]\end{tabular}  \\ \hline
\textsc{Hypertext}      & 113   & 2{,}196       & 20{,}818           & 368    & 20                         \\
\textsc{Workplace}      & 92    & 755         & 9,827            & 214    & 20                         \\
\textsc{Hospital}       & 75    & 1{,}139       & 32{,}424           & 865   & 20                          \\
\textsc{Primary School} & 242   & 8{,}317       & 125{,}773          & 1{,}039   & 20                          \\
\textsc{High School}    & 126   & 1{,}710       & 28{,}561           & 453     & 20                          \\
\textsc{Reality Mining} & 64    & 722         & 13{,}131           & 410 & 5\\
\end{tabular}
\end{ruledtabular}
\end{table*}

\begin{enumerate}
     \item{\emph{Interval shuffling:} For each edge of a network, we uniformly randomly permute the inter-event times, except that we fix the times of the first and last events. This type of shuffling preserves the distribution of inter-event times of each edge. It also preserves the structure of an aggregate time-independent network (including the weight of each edge), which we construct by setting the weight of each edge to be the number of events between its incident nodes. However, interval shuffling destroys the temporal correlations of each edge and across different edges.} 
         \item{\emph{Shuffled time stamps:} We replace time stamps of two random events from different edges of a network. Specifically, we choose two edges uniformly at random, choose one event uniformly at random for each selected edge, and swap the two chosen events. We repeat this procedure for the number of event times in the network. This randomization method preserves the number of events on each edge. It also preserves the set of tevent times of the network, including the multiplicity of each event time. However, shuffling time stamps destroys the distribution of inter-event times of each edge.}
    \item{\emph{Random times:} For each edge of a network, we redistribute the same number of events as the original number independently according to a uniform density on the time window $[0, T]$, where $T$ is the time of the last event in the network. This procedure corresponds approximately to assigning an independent Poisson process to each edge. For each edge, the rate of this process is equal to the number of events on the edge divided by $T$.}
          \item{\emph{Random edge shuffling:} Given a network, we rewire a pair of edges that we choose uniformly at random.
           Specifically, we pick two edges, $(v_i, v_j)$ and $(v_{i'}, v_{j'})$, uniformly at random. We then replace the two edges $(v_i, v_j)$ and $(v_{i'}, v_{j'})$ by the edges $(v_i, v_{j'})$ and $(v_{i'} , v_j)$. This rewiring preserves the time stamps of the two edges. We repeat this procedure for the number of event times in the network. In contrast to the previous three types of randomization, random edge shuffling destroys the structure (by changing the adjacency matrix) of the aggregate time-independent network in which the weight of each edge is the number of events between its incident nodes.}  
     \end{enumerate}



\subsection{Aggregate networks}

We now compare the convergence speeds of the dynamics on tie-decay networks with those in corresponding aggregate networks with the same mean over time of the weight of each edge. The aggregate networks that we consider correspond to {assuming continuous-time Laplacian dynamics of the form $\frac{{\rm d}\bm x}{{\rm d}t} = - \bm x L^{\top}$, where $L$ is the combinatorial Laplacian matrix of the aggregate network. Note that $L$ is constant throughout the time window $[0, T]$.} 
Given a temporal network, we define the weight $w_{ij}$ of edge $e_{ij}$ of the aggregate network by
\begin{equation}
	w_{ij}= \frac{1}{T}\sum_{\ell} \int_{t(i,j,\ell)}^{T} \exp({-\alpha(x-t(i,j,\ell))}) {\rm d}x\,,
\label{eq:aggregate_edge_weight}
\end{equation}
where $t(i,j,\ell)$ is the time of the $\ell$th event on edge $e_{ij}$. Note that $w_{ij}$ is equal to the mean over time of the weight of edge $e_{ij}$ in the tie-decay network. 
{A small decay rate $\alpha$ leads to a large edge weight $w_{ij}$.}


\section{Data Sets}
\label{sec:data_sets}

We construct tie-decay networks from the following six data sets.

\textsc{Hypertext}: This data set was collected during the ACM Hypertext 2009 conference over about 2.5 days \citep{isella2011s, hypertext}.  
We construct a time-dependent network of face-to-face proximity of conference attendees. The data set has a time resolution of $\Delta t = 20$ seconds. This ``time resolution'' is 
the interval between consecutive observations of events. 
Each entry in the data set has the form $(t, i, j)$, where $v_i$ and $v_j$ are the IDs of the people in ``contact'' and $t$ is the time of the contact event. This data set has 113 nodes, $2{,}196$ edges, and $20{,}818$ events. One of these nodes is an almost isolated node that has only one edge (with two events). In Appendix \ref{sec:nearly_isolated}, we study influence of this node on the convergence speed of the opinion dynamics.

\textsc{Workplace}: This data set consists of the contacts between individuals in an office building in France from 24 June to 3 July in 2013 \citep{genois2015data, workplace}. The time resolution of the data set is $\Delta t=20$ seconds. The network has $92$ nodes, $755$ edges, and $9{,}827$ events. 

\textsc{Hospital}: This data set consists of the contacts between individuals, where the individuals are either patients or healthcare workers (HCWs), in a hospital ward in Lyon, France from Monday 6 December 2010 at 1:00 pm to Friday 10 December 2010 at 2:00 pm \citep{hospital, vanhems2013estimating}. The study includes 29 patients and 46 HCWs. The time resolution of the data set is $\Delta t=20$ seconds.
The network has $75$ nodes, $1{,}139$ edges, and $32{,}424$ events. 

\textsc{Primary School}: This data set consists of the contacts between 232 primary-school children and 10 teachers in France \citep{gemmetto2014mitigation, primary, stehle2011high}. The people were recorded for two consecutive days in October 2009. The time resolution of the data set is $\Delta t=20$ seconds.
The network has $242$ nodes, $8{,}317$ edges, and $125{,}773$ events. 

\textsc{High School}: This data set consists of the contacts between students in three classes in a high school in Marseille, France \citep{fournet2014contact, highschool}. The people were recorded for four consecutive days in December 2011. The time resolution of the data set is $\Delta t=20$ seconds.
The network has $126$ nodes, $1{,}710$ edges, and $28{,}561$ events.

\textsc{Reality Mining}: This is a subset of the data that were collected in an experiment that was conducted with students at Massachusetts Institute of Technology over nine months from September 2004 to May 2005 \citep{eagle2006reality}. This subset of the data was also used in Refs.~\cite{masuda2019detecting} and \cite{scholtes2014causality}. The students were given smartphones, and their pairwise proximity was recorded via a Bluetooth channel. The time resolution of the data set is $\Delta t=5$ seconds.
The network has $64$ nodes, $722$ edges, and $13{,}131$ events. 

\medskip

In Table \ref{tab:original_networks}, we summarize the number of nodes, the number of edges, the number of events, the mean number of events per node, and the time resolution of each of these six data sets.



\section{Comparison of the spectral gaps for empirical tie-decay networks, randomized tie-decay networks, and aggregate networks}\label{sec:results_randomized_aggregate}

We numerically examine the convergence speed of continuous-time Laplacian dynamics on tie-decay networks (see Sec.~\ref{sec:laplacian_dynamics}) that we construct from the six empirical data sets (see Sec.~\ref{sec:data_sets}). We compare the convergence speeds of the Laplacian dynamics on empirical tie-decay networks with those of Laplacian dynamics on randomized and aggregate networks (see Sec.~\ref{sec:random_aggregate}).


\begin{figure*}[th]
\includegraphics[width=0.903\textwidth]{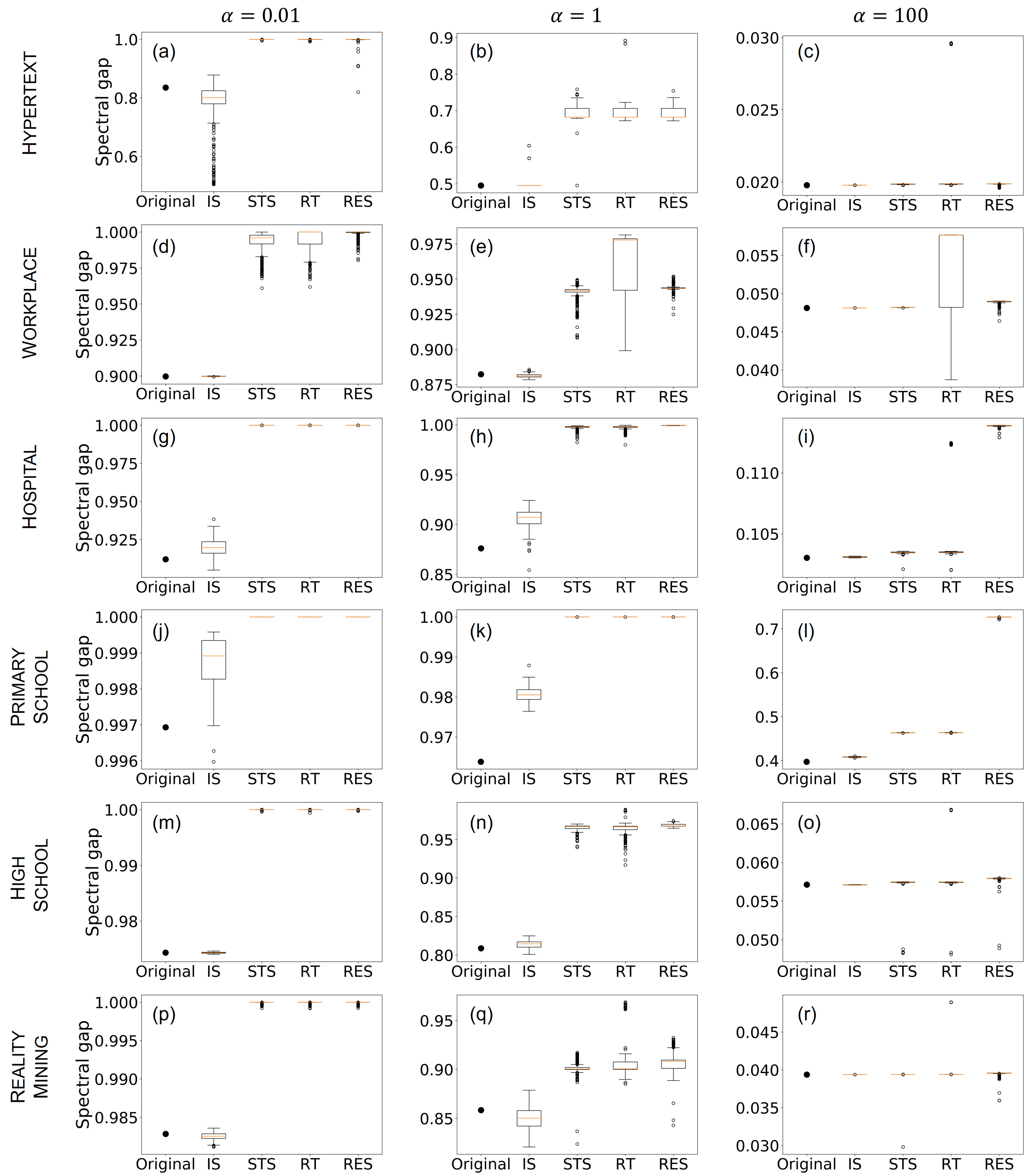}
  \caption{\label{fig:comparison_alpha_all}Comparison of the spectral gaps of $M(T)$ between the original and randomized networks. We examine decay rates of $\alpha=0.01$, $\alpha=1$, and $\alpha=100$. We calculate the spectral gap of $M(T)$ for each of 1{,}000 randomized networks for each type of randomization. The acronyms IS, STS, RT, and RES stand for interval shuffling, shuffled time stamps, random times, and random edge shuffling, respectively. We use box plots to show five-number summaries of the distributions; these quantities are the first quartile ($Q_1$), the median, the third quartile ($Q_3$), the minimum without outliers ($Q_1-1.5 \, \times \, {\rm IQR}$), and the maximum without outliers ($Q_3+1.5 \, \times \, {\rm IQR}$), where ${\rm IQR}=Q_3-Q_1$. Open circles indicate outliers.}
\end{figure*}

\begin{figure*}[t]
\includegraphics[width=\textwidth]{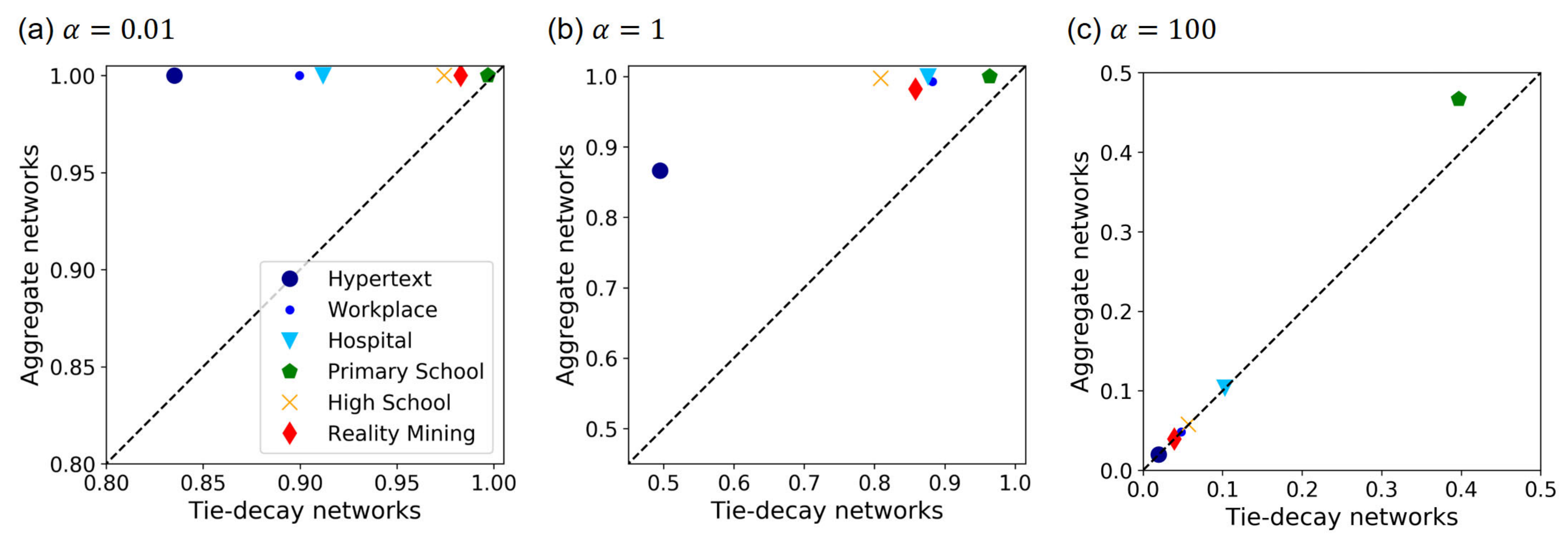}
  \caption{\label{fig:comparison_aggregate}Comparison of the spectral gaps of $M(T)$ for the tie-decay networks to those for the corresponding aggregate networks for (a) $\alpha=0.01$, (b) $\alpha=1$, and (c) $\alpha=100$.}
\end{figure*}

For the continuous-time Laplacian dynamics on tie-decay networks, we quantify the convergence speed of opinion dynamics by the spectral gap \cite{van2010graph} of the matrix $M(t_n)$. The spectral gap of $M(t_n)$ is defined by the difference between the largest-magnitude eigenvalue 
of $M(t_n)$ and the eigenvalue with the second-largest magnitude \cite{van2010graph}. 
The largest-magnitude eigenvalue of  $M(t_n)$ is equal to $1$. Because the eigenvalue $1$ has the corresponding left eigenvector $(1, \ldots, 1)$, which is associated with the nodes reaching a consensus opinion, Eq.~\eqref{eq:x_M} implies that the spectral gap of $M(t_n)$ quantifies the speed at which consensus occurs in continuous-time Laplacian dynamics on tie-decay networks between times $0$ and $t_n$. A large spectral gap implies a fast approach to consensus. For the discrete-time DeGroot model (see Sec.~\ref{sec:Degroot_model}), the speed of convergence is governed by the spectral gap of the product of the $\tilde{B}$ matrices in Eq.~\eqref{eq:DeGroot_5}.

We compare the original networks and randomized networks by calculating the spectral gap of $M(T)$ for decay rates of $\alpha=0.01$, $\alpha=1$, and $\alpha=100$. We show the results for the six data sets in Fig.~\ref{fig:comparison_alpha_all}.
When $\alpha=0.01$, the spectral gaps of $M(T)$ for the randomized networks are significantly larger than those for the original networks in all cases except for interval shuffling applied to the \textsc{Hypertext}, \textsc{Workplace}, \textsc{Hospital}, \textsc{High School}, and \textsc{Reality Mining} data sets (see Figs.~\ref{fig:comparison_alpha_all}(a,d,g,j,m,p)). When $\alpha=1$, the spectral gaps of $M(T)$ for the randomized networks are significantly larger than those for the original networks in all cases except for interval shuffling applied to the \textsc{Hypertext}, \textsc{Workplace}, \textsc{High School}, and \textsc{Reality Mining} data sets (see Figs.~\ref{fig:comparison_alpha_all}(b,e,h,k,n,q)). 
When $\alpha=100$, the spectral gaps of $M(T)$ for the randomized networks are significantly larger than those for the original networks in all cases except for interval shuffling applied to the \textsc{Hypertext}, \textsc{Workplace}, \textsc{Hospital}, \textsc{High School}, and \textsc{Reality Mining} data sets and random times applied to the \textsc{Workplace} data set (see Figs.~\ref{fig:comparison_alpha_all}(c,f,i,l,o,r)). Therefore, in a majority of the examined cases and for all three values of $\alpha$, the randomization increases the spectral gap of $M(T)$ and thus yields faster convergence. We do not observe any situations in which the spectral gap of $M(T)$ for an original network is significantly larger than that for an associated randomized network. {Figure~\ref{fig:comparison_alpha_all} also leads to several other observations. First, the} spectral gaps of $M(T)$ for the randomized networks that we obtain from random edge shuffling tend to be larger than those for the other types of randomization. For example, see the results for the randomizations of the \textsc{Hospital} and \textsc{Primary School} data sets for $\alpha=100$. {Second,} most of the cases in which randomization does not significantly increase the spectral gap of $M(T)$ occur when we use the interval-shuffling method of randomization. {Third, the networks in which interval shuffling increases the spectral gap are those with the largest number of events per node (specifically, \textsc{Primary School} for $\alpha = 0.01$, \textsc{Primary School} and \textsc{Hospital} for $\alpha = 1$, and \textsc{Primary School} for $\alpha = 100$). Fourth, as we discuss further in Section~\ref{sec:conclusions}, the differences in the spectral gaps of $M(T)$ between the original and randomized networks and between the different types of randomized networks tend to be small for large decay rates.}


{We interpret the above results in Fig.~\ref{fig:comparison_alpha_all} as follows. Overall, the spectral gaps after randomizing using random times tend to be larger than those after randomization using interval shuffling. This result suggests that the deviations of the distributions of the inter-event times from exponential distributions decelerates the convergence (see Table~\ref{tab:four_randomization}). Similarly, the spectral gaps after randomizing using random edge shuffling tend to be larger than those after applying the other randomization methods. Random edge shuffling destroys the structure of an aggregate network, but the other randomization methods do not.
This result suggests that the particular structures of the aggregations of the original networks lead to slower convergence
than in randomized aggregate networks that preserve the weighted degree of each node.}


We now compare the spectral gaps of $M(T)$ for the tie-decay networks to those for their corresponding aggregate networks. 
We set the weight of each edge {in an aggregate network} to be equal to the mean of the weights of that edge in the associated tie-decay network over all times in the window $[0, T]$. 
In Fig.~\ref{fig:comparison_aggregate}, we show the spectral gaps of $M(T)$ for the tie-decay networks and their corresponding aggregate networks for decay rates of $\alpha=0.01$, $\alpha=1$, and $\alpha=100$ for the six data sets.
{The spectral gap is larger for progressively smaller decay rates $\alpha$. This occurs because a small decay rate results in large edge weights 
(see Eq.~\eqref{eq:aggregate_edge_weight}).}
The figure suggests that, for all of our data sets and all three of these decay rates, the spectral gaps of $M(T)$ for the aggregate networks are larger than those for the corresponding original
networks. 
{We also observe that the spectral gaps for the tie-decay networks that we construct from the \textsc{Hypertext} data set are much smaller than those for the other data sets. This difference appears to arise from the almost isolated node in the \textsc{Hypertext} network. Indeed, as we will show in Appendix~\ref{sec:nearly_isolated}, when we exclude this node, the spectral gaps for the resulting tie-decay networks are larger than $0.99$ for $\alpha=0.01$ and $\alpha=1$. These results are similar to those for the other data sets.}

Our results for the aggregate networks are consistent with those for the randomized temporal networks. This makes intuitive sense because (like in the randomized temporal networks) our aggregations of the networks destroy the temporal structures of the original networks. Recall that interval shuffling, shuffled time stamps, and random times (which destroy the temporal structures of the event sequences in the original data sets to different extents) lead to spectral gaps of $M(T)$ that are larger than those in the associated original temporal
networks.


\section{Non-monotonicity of the spectral gap}\label{sec:non_monotonicity}


\subsection{Spectral gap as a function of the decay rate $\alpha$}
\label{sec:result_A}

\begin{figure}[t]
\includegraphics[width=0.49\textwidth]{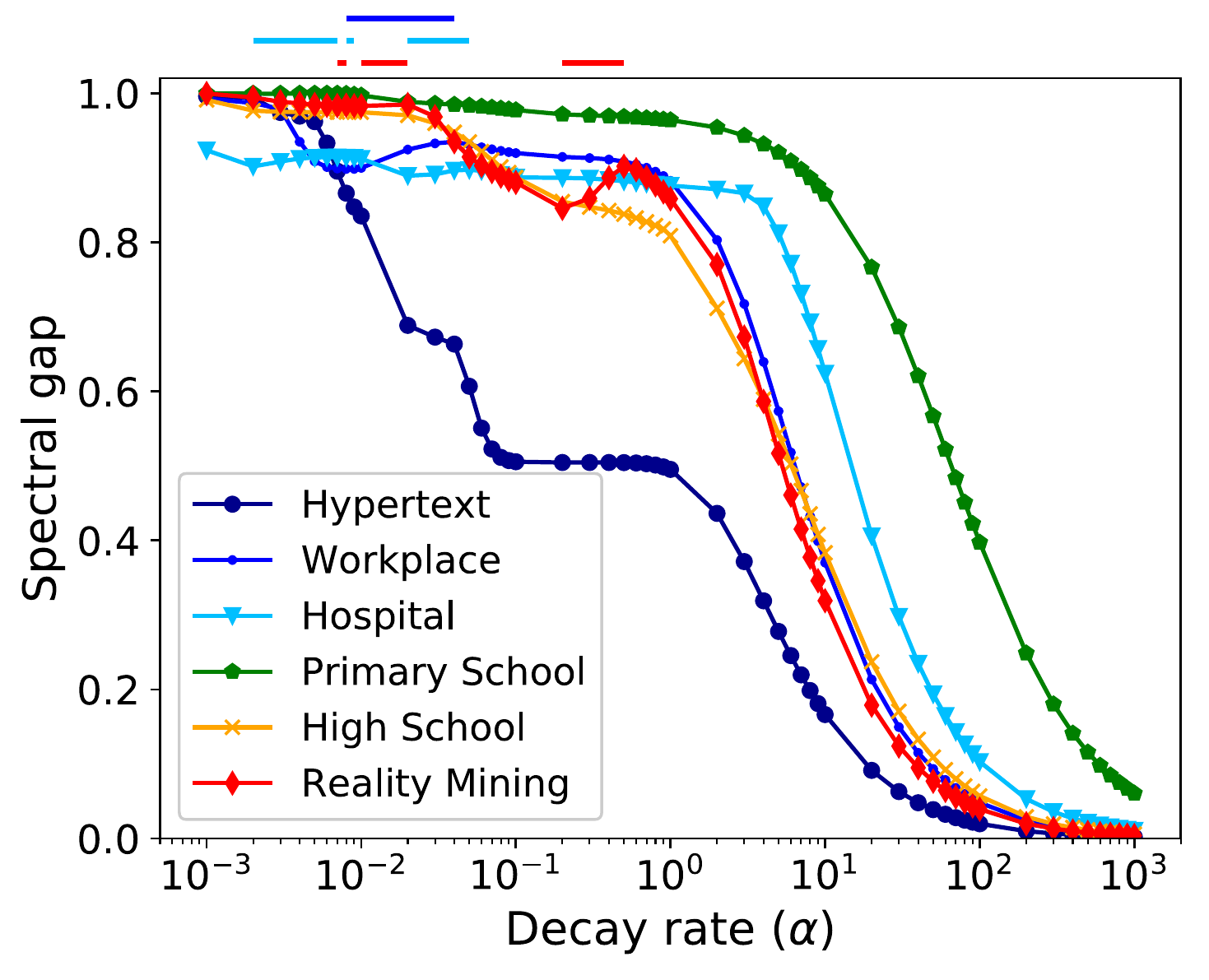}
\caption{\label{fig:comparison_six_data_sets} Comparison of the spectral gaps of $M(T)$ for the six data sets for a range of values of the decay rate $\alpha$. {The horizontal line segments above the main figure indicate the values of $\alpha$ for which the slopes of the curves are positive for the \textsc{Workplace}, \textsc{Hospital}, and \textsc{Reality Mining} data sets.} 
}
\end{figure}

In Fig.~\ref{fig:comparison_six_data_sets}, we show the spectral gaps of $M(T)$ for tie-decay networks that we construct from the aforementioned six data sets. {In accordance with our results in Fig.~\ref{fig:comparison_aggregate}, the spectral gaps for \textsc{Hypertext} are much smaller than those for the other data sets.} 
From {Fig.~\ref{fig:comparison_six_data_sets}}, we see that a larger value of $\alpha$ (i.e., faster tie decay) tends to lead to a smaller spectral gap of $M(T)$ and hence to slower convergence. This makes intuitive sense because a large decay rate implies that the ties between nodes weaken more rapidly. Additionally, in three of the six data sets (specifically, \textsc{Workplace}, \textsc{Hospital}, and \textsc{Reality Mining}), the spectral gaps of $M(T)$ do not decrease monotonically with $\alpha$. This is counterintuitive, so it is worth further exploration. 

As a quick check, we calculate the spectral gap of each factor of $M(T)$ as a function of the decay rate $\alpha$. Recall that such a factor takes the form
\begin{equation}\label{factor}
	\exp\left[\frac{\tilde{L}({t_{n-1}^+})^{\top}}{\alpha}(e^{-\alpha (t_n-t_{n-1})}-1)\right]\,, \quad n \in \{1, 2, \ldots\}\,.
\end{equation}
We find that the spectral gap of each factor \eqref{factor} of $M(T)$ decreases monotonically with $\alpha$ for all of the data sets. This makes intuitive sense, so there is nothing ``strange'' at the level of any single interval between event times.
 Therefore, it seems that the product of the factors of $M(T)$ leads to the above counterintuitive result for three data sets. 
In Appendix~\ref{sec:randomization_monotonicity}, we study the spectral gap of $M(T)$ as a function of $\alpha$ for the randomized networks for these three data sets (specifically, \textsc{Workplace}, \textsc{Hospital}, and \textsc{Reality Mining}). 
We observe non-monotonic dependence on $\alpha$ when we apply interval shuffling to each of these three data sets, but we do not observe such non-monotonicity in these data sets when we use the other three randomization methods.

\subsection{Spectral gap as a function of time}
\label{sec:SG_time}

\begin{figure*}[t]
\includegraphics[width=\textwidth]{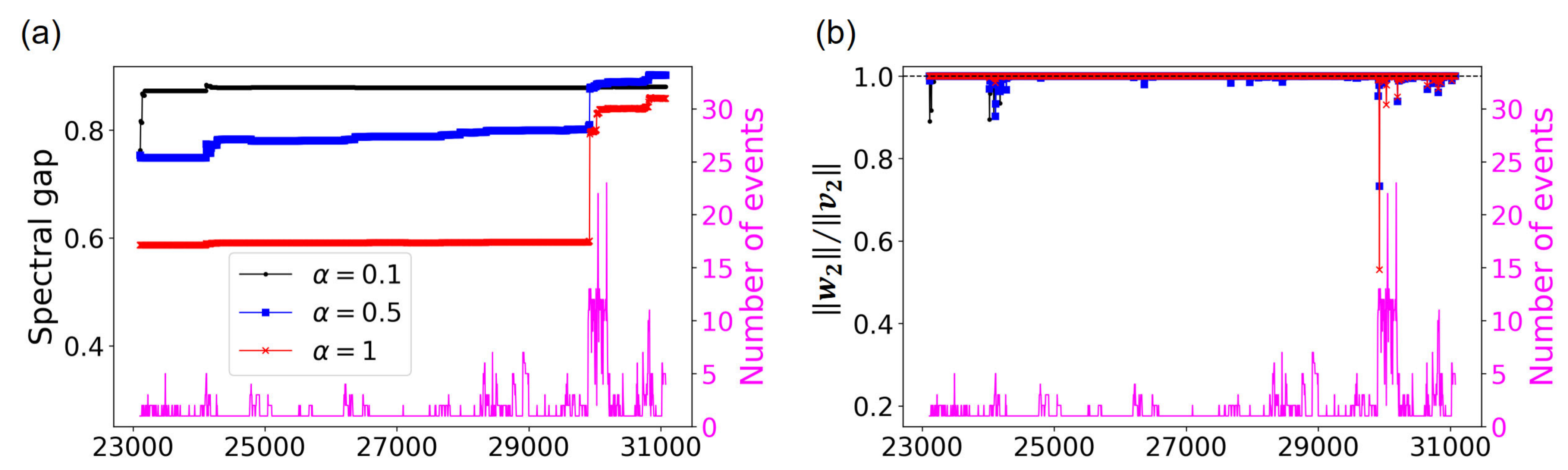}
\caption{\label{fig:RealityMining} The spectral gap of $M(t_n)$ for the \textsc{Reality Mining} data set as a function of time. We also show the number of events at each time. (a) The spectral gap. (b) The ratio ${\|\bm w_2\|}/{\|\bm v_2\|}$, which indicates how much {the time-independent networks that we construct from the events at each time} shrink the length of the normalized Fiedler vector of the matrix that encodes the original tie-decay network.
}
\end{figure*}

To help understand the non-monotonicity of the spectral gap of $M(T)$ as a function of the decay rate $\alpha$, we study the spectral gap as a function of time.
In Fig.~\ref{fig:comparison_six_data_sets}, we observed such non-monotonicity in three data sets (\textsc{Workplace}, \textsc{Hospital}, and \textsc{Reality Mining}). Among these data sets, the \textsc{Reality Mining} data set has the most prominent non-monotonicity, which occurs
between $\alpha \approx 0.1$ and $\alpha \approx 1$. Therefore, we focus on three $\alpha$ values in the interval $[0.1, 1]$. We show the spectral gap of $M(t_n)$ for the \textsc{Reality Mining} data set as a function of time for $\alpha = 0.1$, $\alpha = 0.5$, and $\alpha = 1$ in Fig.~\ref{fig:RealityMining}(a). {In this figure, we also show the number of events at each time.} We observe that the magnitude of the change in the spectral gap at each time does not depend monotonically on the number of events. Additionally, the spectral gap of $M(t_n)$ need not increase monotonically with time. Moreover, the dynamics of the spectral gap of $M(t_n)$ depends on the decay rate $\alpha$. 
The most drastic change in the spectral gap of $M(t_n)$ for $\alpha=0.5$ and $\alpha = 1$ occurs at $t_n=29{,}925$. At this time, the increase in the spectral gap of $M(t_n)$ is largest for $\alpha=1$, second largest for $\alpha=0.5$, and smallest for $\alpha=0.1$. In fact, for $\alpha=0.1$, the spectral gap of $M(t_n)$ changes little after $t_n = 29{,}925$, which is why the spectral gap of $M(t_n)$ for $\alpha=0.5$ exceeds that for $\alpha=0.1$ after this time, whereas the spectral gap of $M(t_n)$ for $\alpha=0.1$ is larger than that for $\alpha=1$. 
This results in the non-monotonicity of the spectral gap of $M(T)$ that we observed in Fig.~\ref{fig:comparison_six_data_sets}.

\begin{figure*}[tb]
\includegraphics[width=\textwidth]{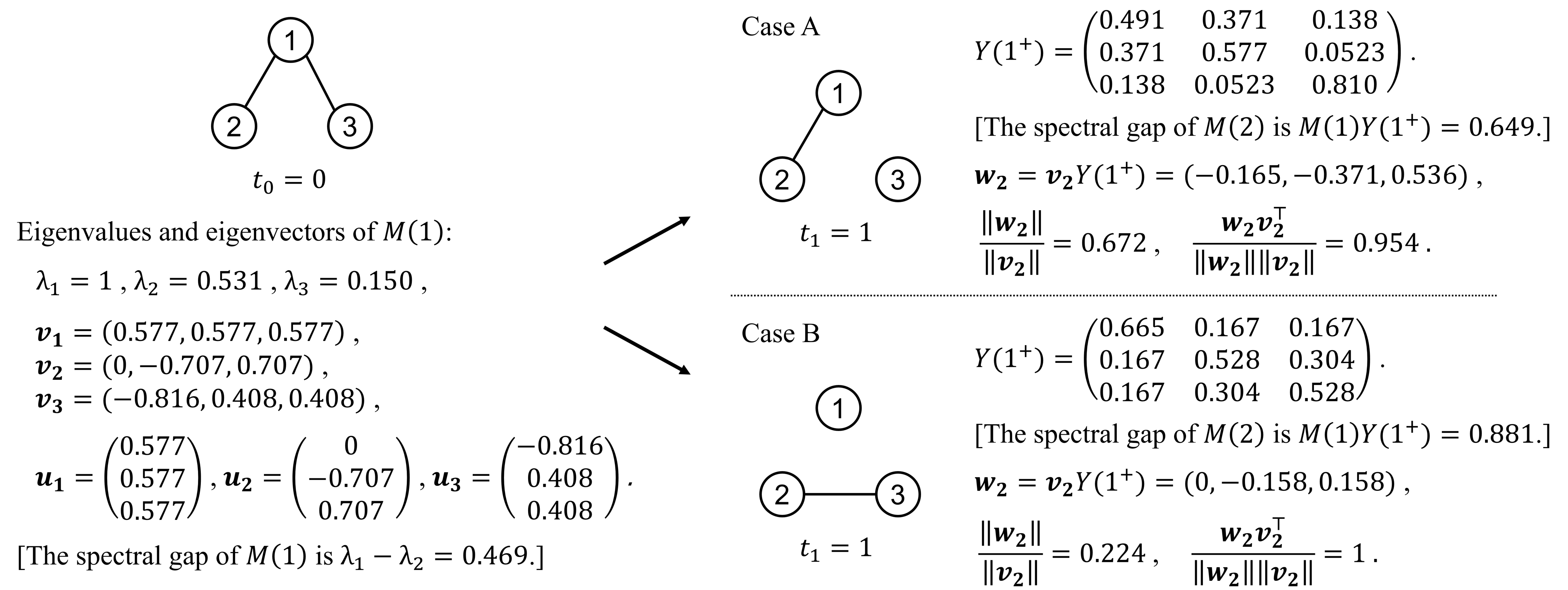}
\caption{Two temporal networks that have the same set of events at time $t = t_0$ but different sets of events at time $t = t_1$. We show the eigenvalues and eigenvectors of the matrix $M(1)$ on the left. Because the two cases, which we denote by A and B, have different events at time $t_1=1$, they
 have different matrices $Y({1^+})$, different spectral gaps of $M(2)$, and different values of ${\|\bm w_2\|}/{\|\bm v_2\|}$.
We show the values of our computations to three digits of precision.}
\label{fig:3_3} 
\end{figure*}
    

\subsection{Updating the Fiedler vector of $M(t_n)$}
\label{sec:Fiedler}

In Fig.~\ref{fig:RealityMining}(a), it seems that the spectral gap of $M(t_n)$ experiences a sudden increase with respect to time. We hypothesize that such a sudden increase occurs between time $t_n$ and time $t_{n+1}$ when the time-independent network that we construct from the events at time $t_n$ is effective at shrinking the length of the normalized Fiedler vector (i.e., the eigenvector that is associated with the eigenvalue with the second-largest magnitude) of the matrix $M(t_n)$ (where we use an arbitrary 
norm to calculate the length). Note that $M(t_n)$ indicates the matrix just before the events at time $t_n$. 
To try to explain this hypothesis, we reexamine the matrix $M(t_n)$ and its spectral gap. 

We rewrite Eq.~\eqref{eq:finite_product} as
\begin{equation}
	M(t_{n+1}) = M(t_{n})Y({t_n^+})\,, \quad n \geq 1\,,
\label{eq:M_X_Y}
\end{equation}
and we define the square matrix $Y({t_{n}^+})$ by
\begin{equation}
	Y({t_n^+})=\exp\left[\frac{\tilde{L}({t_{n}^+})^{\top}}{\alpha}(e^{-\alpha (t_{n+1}-t_{n})}-1)\right]\,. 
\label{eq:Y}
\end{equation}

Under the assumption of diagonalizability, we decompose the matrix $M(t_{n})$ as follows:
\begin{equation}
	M(t_{n})=\sum_{i=1}^{N}{\lambda_i\bm u_i \bm v_i}\,,
\label{eq:X_decomposition}
\end{equation}
where $\lambda_i$ is the $i$th eigenvalue of $M(t_{n})$ in descending order in magnitude, $\bm u_i$ is the associated right eigenvector, and $\bm v_i$ is the associated left eigenvector. Note that $\lambda_1 = 1$ and ${\bm v}_1 = (1, \ldots, 1)$. We normalize the eigenvectors such that $\bm v_i \bm u_j = \delta_{ij}$, where $\delta_{ij}$ is the Kronecker delta. Let $\bm v_2$ denote the left Fiedler vector of $M(t_{n})$. In general, we do not expect the matrices $M(t_{n})$ and $Y({t_{n}^+})$ to have the same eigenspace. However, to explain the idea that underlies our hypothesis, we assume for now the unrealistic situation in which $Y({t_{n}^+})$ and $M(t_{n})$ have the same eigenspace, such that their eigenvectors are the same. Let $\mu_i$ denote the eigenvalue of $Y({t_{n}^+})$ with corresponding right eigenvector $\bm u_i$ and left eigenvector $\bm v_i$. We write
\begin{equation}
	Y({t_{n}^+}) = \sum_{i=1}^N \mu_i \bm u_i \bm v_i\,,
\label{eq:Y_new}
\end{equation}
and we note that $\mu_1 = 1$. Using Eqs.~\eqref{eq:M_X_Y}, \eqref{eq:X_decomposition}, and \eqref{eq:Y_new}, we obtain
\begin{equation}
	M(t_{n+1}) = M(t_{n})Y({t_{n}^+}) = \sum_{i=1}^N \lambda_i \mu_i \bm u_i \bm v_i\,.
\label{eq:XY_revised}
\end{equation}
Equation~\eqref{eq:XY_revised} indicates that the spectral gap of $M(t_{n+1})$ is $1- \min_{2\le i\le N} \{ \lambda_i \mu_i \}$.
Recall that $|\lambda_2| \ge \cdots \ge |\lambda_N|$. For a given $M(t_{n+1})$, the spectral gap of $M(t_{n+1})$ tends to be large if $|\lambda_2 \mu_2|$ is small, which tends to be the case when $|\mu_2|$ is small. Because Eq.~\eqref{eq:Y_new} implies that $\bm v_2 Y({t_{n}^+}) = \mu_2 \bm v_2$, the length of the Fiedler vector of $M(t_{n})$ decreases by a large amount as a result of multiplication by $Y({t_{n}^+})$ if $|\mu_2|$ is small. Therefore, if $|\mu_2|$ is small, the spectral gap of $M(t_{n+1})$ tends to be large. This explains our hypothesis {that a sudden increase in the spectral gap of $M(t_n)$ occurs when the time-independent network that we construct from the events at time $t_n$ is effective at shrinking the length of the normalized Fiedler vector of $M(t_n)$}.

In general, $M(t_{n})$ and $Y({t_{n}^+})$ have different eigenspaces. For diagonalizable $Y({t_{n}^+})$, we write
\begin{equation}
	Y({t_{n}^+})=\sum_{i=1}^{N}{\bm u_i \bm w_i}\,,
\label{eq:Y_decomposition}
\end{equation}
where $\bm w_i$ is an $N$-dimensional row vector. The $j$th element of $\bm w_i$ is the coefficient of $\bm u_i$ when we express the $j$th column of $Y({t_{n}^+})$ as a linear combination of $\bm u_1$,\ldots,$\bm u_N$. Equation \eqref{eq:Y_decomposition} indicates that 
$\bm w_i=\mu_i \bm v_i$ if $M(t_{n})$ and $Y({t_{n}^+})$ have the same eigenspace.
Using Eq.~\eqref{eq:Y_decomposition}, we obtain
\begin{equation}
	\bm v_2 Y({t_{n}^+})=\bm w_2
\label{eq:v_2_w_2}
\end{equation}
because $\bm v_i\bm u_j=\delta_{ij}$. Therefore, the length of $\bm w_2$ represents how much the normalized Fiedler vector $\bm v_2$ shrinks (or expands) in length by right-multiplying by the matrix $Y({t_{n}^+})$. Because $M(t_{n})$ and $Y({t_{n}^+})$ have different eigenspaces in general, it is not true in general that $\bm w_2\propto\bm v_2$. However, if the length of $\bm w_2$ is small, the situation is analogous to having a small value of $|\mu_2|$ when $M(t_{n})$ and $Y({t_{n}^+})$ have the same eigenspace. Therefore, 
we expect the spectral gap of $M(t_{n})$ to increase by a large amount as a result of right-multiplication of $M(t_{n})$ by $Y({t_{n}^+})$ if the length of $\bm w_2$ is small. Based on this reasoning, we measure ${\|\bm w_2\|}/{\|\bm v_2\|}$ as a function of time, where $\| \cdot \|$ is the 2-norm. 

We show $\|\bm w_2\| / \| \bm v_2 \|$ for the \textsc{Reality Mining} data set for several values of the decay rate ($\alpha = 0.1$, $\alpha = 0.5$, and $\alpha = 1$) in Fig.~\ref{fig:RealityMining}(b). The figure indicates that $\|\bm w_2\| / \| \bm v_2 \|$ depends on $\alpha$ in qualitatively different manners at different times $t_n$. For example, the decrease in ${\|\bm w_2\|}/{\|\bm v_2\|}$ at $t_n=29{,}925$ is largest for $\alpha=1$, second largest for $\alpha=0.5$, and smallest for $\alpha=0.1$. By contrast, the decrease in ${\|\bm w_2\|}/{\|\bm v_2\|}$ at $t_n=24{,}020$ is largest for $\alpha=0.1$, second largest for $\alpha=0.5$, and smallest for $\alpha=1$. The decrease in ${\|\bm w_2\|}/{\|\bm v_2\|}$ at $t_n=24{,}115$ is largest for $\alpha=0.5$. These results are consistent with the dependence of the spectral gap on $\alpha$ and $t$ in Fig.~\ref{fig:RealityMining}(a). {In other words, the spectral gap of $M(t_n)$ in Fig.~\ref{fig:RealityMining}(a) tends to increase when the ratio ${\|\bm w_2\|}/{\|\bm v_2\|}$ is small in Fig.~\ref{fig:RealityMining}(b).} We obtain the same qualitative results for the \textsc{Workplace} and \textsc{Hospital} data sets (see Appendix \ref{sec:appendix_non_monotonic}).  

To develop intuition, we show an illustrative example of a computation of ${\|\bm w_2\|}/{\|\bm v_2\|}$ in Fig.~\ref{fig:3_3}. In this example, we consider two tie-decay
networks with three nodes. These two networks have the same set of events at time $t_0 = 0$ but different sets of events at time $t_1 = 1$. Therefore, the tie-decay networks of the two cases --- which we label as case A and case B --- are the same for $t \in [0,1)$ but different for $t \ge 1$. We set the decay rate to $\alpha=1$. Consider the spectral gap of $M(2)$, which is equal to $M(1)Y({1^+})$ by Eq.~\eqref{eq:M_X_Y}. The two tie-decay networks have the same matrix $M(1)$, but the matrix $Y({1^+})$ is different for the two networks. The spectral gap of $M(2)$ is larger in case B than it is in case A. In accordance with this observation, the ratio ${\|\bm w_2\|}/{\|\bm v_2\|}$ is smaller in case B than it is in case A. Consequently, we see that the event at time $t_1$ is more effective at shrinking the length of the normalized Fiedler vector $\bm v_2$ and more effective at increasing the spectral gap of $M(2)$ in case B than it is in case A. Note that $\bm w_2$ is not parallel to $\bm v_2$ in case A, because ${\bm w_2}{\bm v_2}^\top/({\|\bm w_2\|}{\|\bm v_2\|}) \approx 0.954$, which indicates that $M(1)$ and $Y({1^+})$ do not have the same eigenspace in case A.


\section{Conclusions and Discussion}
\label{sec:conclusions}

We studied opinion dynamics on tie-decay networks. Specifically, we formulated continuous-time Laplacican dynamics and a discrete-time DeGroot model of opinion dynamics on tie-decay networks. Using numerical computations, we {examined} the convergence speeds of continuous-time Laplacian dynamics on tie-decay networks that we constructed from empirical social-contact data by calculating the spectral gap of the matrix $M(t_n)$, which maps the initial opinions of the nodes of a network to the opinions of the nodes at time $t_n$.

The randomization methods that we considered often increased the spectral gaps for the temporal networks that we examined. We also observed that 
aggregate networks always have larger spectral gaps than their corresponding empirical tie-decay networks. These results are consistent with previous studies that illustrated that spreading dynamics can be {slower on temporal networks than on corresponding aggregate networks} \cite{fernandez2011update, holme2012temporal,iribarren2009impact, karsai2011small, masuda2013temporal,takaguchi2013bursty, vazquez2007impact}.

We examined the influence of the tie-decay rate on the spectral gap of $M(t_n)$. Intuitively, one expects a decrease in the decay rate to accelerate the convergence speed because the influence of each event lasts longer when the decay rate is smaller. However, for some of the data sets, the spectral gap of $M(t_n)$ does not decrease monotonically with respect to the decay rate. We also examined the evolution of the spectral gap in time. As time progresses and more interactions occur, one expects intuitively that the spectral gap will increase. However, we showed empirically that the spectral gap need not increase monotonically with respect to time. 

We showed that the decay rate of a tie-decay network affects the shrinkage of the length of the Fiedler vector of $M(t_n)$. We observed that this dependence was often non-monotonic with respect to the decay rate and that it is consistent with the behavior of the spectral gap of $M(t_n)$.

{When the decay rate is large, we found that the spectral gap is (1) similar for the original networks and their associated random networks and (2) similar for different types of randomized networks. 
(See Fig.~\ref{fig:without_nearlyisolatednode}(c) and the right panels of Fig.~\ref{fig:comparison_alpha_all}.)
When the decay rate is large, the effect of each event tends to be tiny by the time the next event occurs on the same edge. Therefore, it is reasonable to
assume for large decay rates that different events on the same edge act independently to drive opinion dynamics. However, even in this situation,
the order of the events impacts the value of the spectral gap of the combinatorial Laplacian matrix \cite{masuda2013temporal,masuda2016accelerating}. In our investigation,
we saw that different randomizations induce different orderings of the events. However, we did not observe a notable dependency of the spectral gap on different randomizations of the same temporal network. Therefore, the lack of influence from
the previous events when the next event occurs on the same edge is probably not a key reason that the spectral gaps tend to be similar for the original networks and the different randomized networks for large decay rates $\alpha$. 
We do not have a firm understanding of this observation. 
}

There are many interesting future directions that build on our results. {First}, it is desirable to examine how heterogeneous decay rates for different edges influence the convergence speed of opinion dynamics. 
{{Second, it is important to fit} decay rates to empirical data to study the time scales of the influence of events.} {Third,} functional forms of tie-decay dynamics other than ones that are exponential {are also worth examining. Fourth, it is worth considering different rules 
{for changes in tie strengths in response to 
events} and the impact of such rules on
opinion dynamics. 
For example, one can study what happens when
each event between two nodes not only strengthens their mutual relationship but also weakens their relationships with other nodes \cite{gelardi2021from}.} {{Fifth, it is worth studying opinion dynamics on coevolving networks \cite{gross2008adaptive} using the framework of tie-decay networks. To construct such a model, one needs a mechanism (e.g., perhaps an adaptation of the one in \cite{brooks2020}) that generates
discrete events in a manner that depends on the nodes' opinions.}} {Sixth, the} observed non-monotonicity of the spectral gap of $M(t_n)$ may be related to community structure in temporal networks, and exploring this idea may yield insights into bottlenecks of opinion dynamics between communities. Additionally, it is important to investigate whether this non-monotonicity is a property of many dynamical processes or instead results from some peculiarity of the dynamics that we investigated. More generally, we have considered a rather specific form of opinion dynamics, and it is important to investigate the behavior of other types of opinion models (and other types of spreading dynamics \cite{chen2020epidemic}) on tie-decay networks.


\begin{acknowledgments}

KS acknowledges support from the Japan Society for the Promotion of Science
(under Overseas Research Fellowships and Grant No. 19K23531). MAP was supported by the National Science Foundation (under Grant No. 1922952) through the Algorithms for Threat Detection (ATD) program. NM acknowledges support from AFOSR European Office (under Grant No. FA9550-19-1-7024), the Sumitomo Foundation, and the Nakatani Foundation.

\end{acknowledgments}


\appendix

\section{Influence of the almost isolated node in the \textsc{Hypertext} data set}
\label{sec:nearly_isolated}

 \begin{figure*}[ht]
\includegraphics[width=\textwidth]{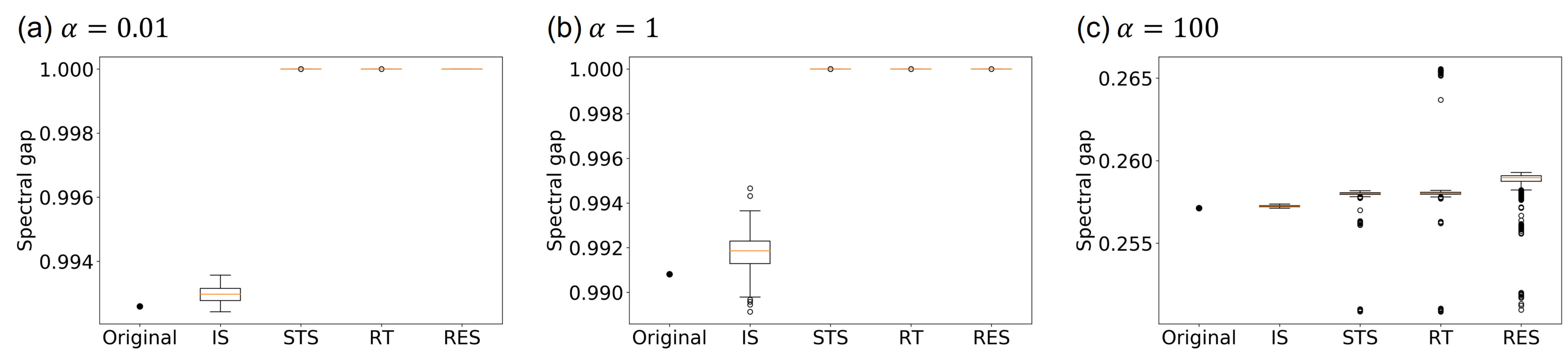}
  \caption{\label{fig:without_nearlyisolatednode}Comparison between the original and randomized networks of the spectral gap of $M(T)$ for the \textsc{Hypertext} network without the almost isolated node. We use decay rates of (a) $\alpha=0.01$, (b) $\alpha=1$, and (c) $\alpha=100$.}
\end{figure*}

 \begin{figure*}[bht]
\includegraphics[width=0.74\textwidth]{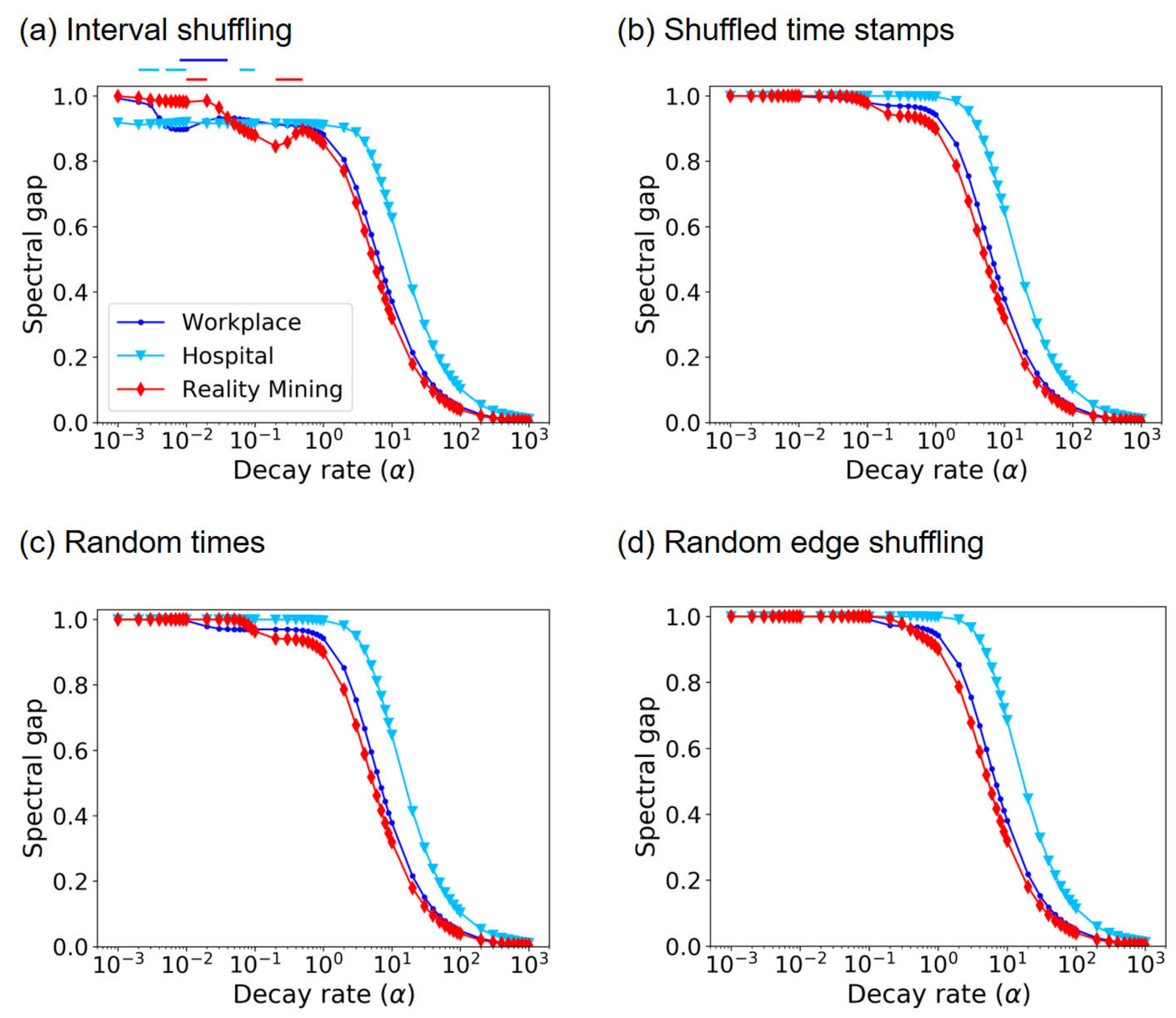}
  \caption{\label{fig:randomization_monotonicity} Comparison of the spectral gaps of $M(T)$ for a range of values of $\alpha$, the four types of randomized networks, and three data sets (\textsc{Workplace}, \textsc{Hospital}, and \textsc{Reality Mining}). We show our results for randomized networks that we produce using (a) interval shuffling, (b) shuffled time stamps, (c) random times, and (d) random edge shuffling. The horizontal line segments above the main figure {in panel (a)} indicate the values of $\alpha$ for which the slopes of the curves are positive for the three data sets. {In panels (b)--(d), the spectral gap is monotonically nonincreasing in all cases.}
  }
\end{figure*}

As we indicated in Sec.~\ref{sec:data_sets}, the \textsc{Hypertext} data set has 113 nodes. This includes one almost isolated node that has only one edge (with two events). In this appendix, we study the influence of the almost isolated node on the spectral gap of $M(T)$ by excluding it
and the two events that are associated with it. In Fig.~\ref{fig:without_nearlyisolatednode}, we show the spectral gap of $M(T)$ when we construct tie-decay networks without the almost isolated node and the two associated events. We again use decay rates of $\alpha=0.01$, $\alpha=1$, and $\alpha=100$. In Figs.~\ref{fig:comparison_alpha_all}(a,b,c), we showed the spectral gap of $M(T)$ when the almost isolated node and the two associated events are present. By comparing Figs.~\ref{fig:comparison_alpha_all}(a,b,c) with Fig.~\ref{fig:without_nearlyisolatednode}, we see that the spectral gaps of $M(T)$ for the networks that include the almost isolated node are smaller than those for the corresponding networks that exclude that node. (See the ``Original'' labels in the figures.) This indicates that the almost isolated node is a bottleneck, as it decelerates the Laplacian dynamics on this network.  

\begin{figure*}[t]
\includegraphics[width=0.95\textwidth]{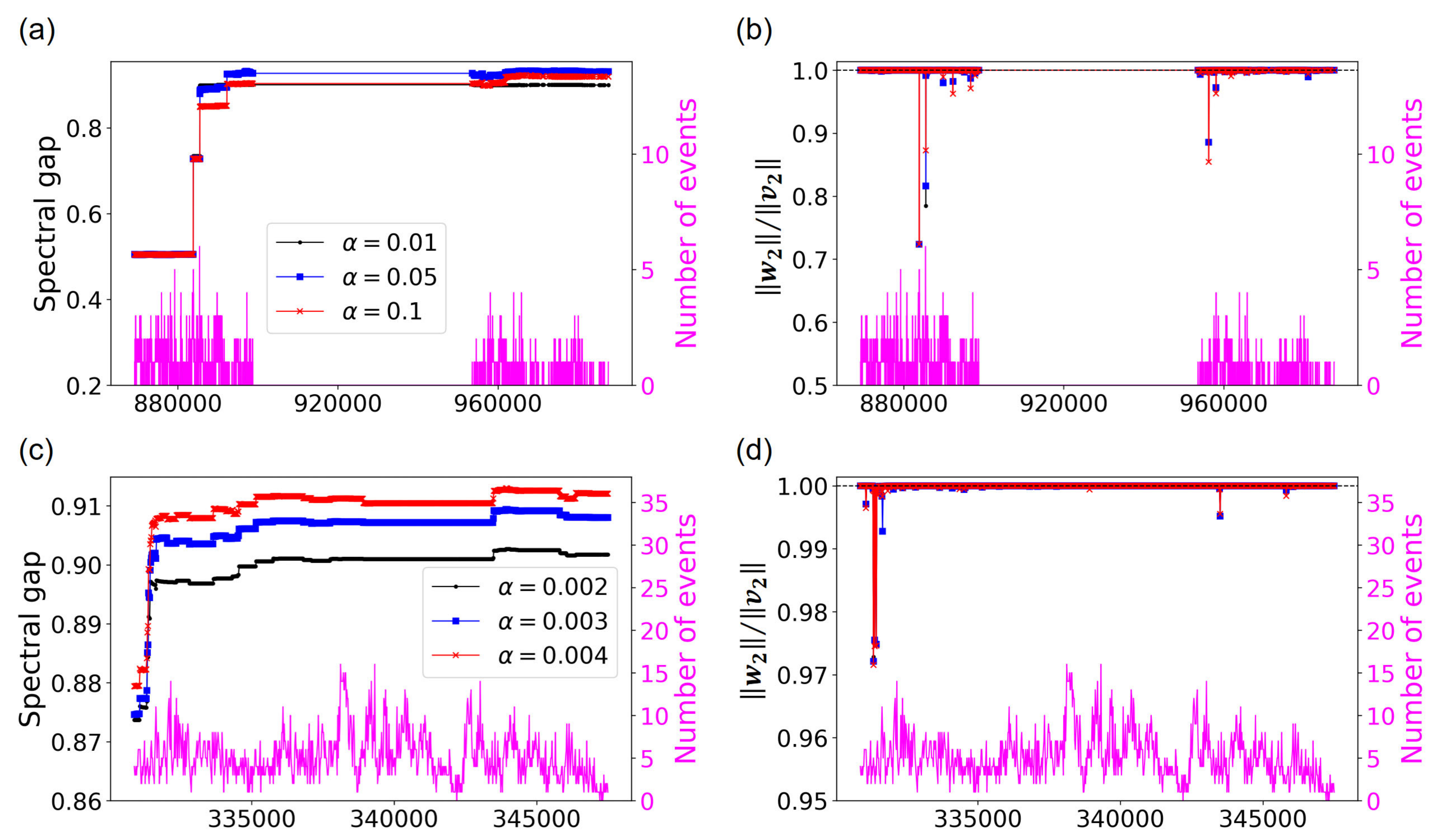}
  \caption{\label{fig:Workplace_hospital}The spectral gaps of $M(t_n)$ for the \textsc{Workplace} and \textsc{Hospital} networks as a function of time. We also show the number of events at each time. (a) The spectral gap and (b) the ratio ${\|\bm w_2\|}/{\|\bm v_2\|}$ for the \textsc{Workplace} network. (c) The spectral gap and (d) the ratio ${\|\bm w_2\|}/{\|\bm v_2\|}$ for the \textsc{Hospital} network.}
\end{figure*}

When we exclude the almost isolated node, the spectral gaps of $M(T)$ for the randomized networks are significantly larger than those for the corresponding original networks for all three of the above decay rates and for all types of randomizations except interval shuffling (see Fig.~\ref{fig:without_nearlyisolatednode}). These results are qualitatively the same as those for the networks that include the almost isolated node (see Fig.~\ref{fig:comparison_alpha_all}).

\section{The spectral gap as a function of decay rate for the randomized networks}\label{sec:randomization_monotonicity}

In Sec.~\ref{sec:non_monotonicity}, we observed non-monotonicity in the spectral gap of $M(T)$ as a function of the decay rate $\alpha$ for three data sets (\textsc{Workplace}, \textsc{Hospital}, and \textsc{Reality Mining}). In this appendix, we study the spectral gap of $M(T)$ as a function of decay rate for the randomized networks that we construct from these data sets.

We show our results for the four randomization methods in Fig.~\ref{fig:randomization_monotonicity}. For each of these data sets, we observe non-monotonicity in the randomized networks that we produce using interval shuffling (see Fig.~\ref{fig:randomization_monotonicity}(a)), but we do not observe it for the networks that we obtain from the other three randomization methods (see Figs.~\ref{fig:randomization_monotonicity}(b)--\ref{fig:randomization_monotonicity}(d)). The non-monotonic behavior in Fig.~\ref{fig:randomization_monotonicity}(a) is similar to what we observed for the original networks in Fig.~\ref{fig:comparison_six_data_sets}. Such non-monotonicity is consistent with our results in Fig.~\ref{fig:comparison_alpha_all}, in which the spectral gaps of $M(T)$ for the randomized networks that we obtained from interval shuffling do not differ significantly from those for the original networks in most cases. 


\section{Non-monotonicity in the \textsc{Workplace} and \textsc{Hospital data sets}}
\label{sec:appendix_non_monotonic}

In this appendix, we study the non-monotonic behavior of the spectral gap of $M(T)$ with respect to the decay rate $\alpha$ in the \textsc{Workplace} and \textsc{Hospital} data sets.

From Fig.~\ref{fig:comparison_six_data_sets}, we see that the spectral gap of $M(T)$ for the \textsc{Workplace} network changes non-monotonically with respect to $\alpha$ when $\alpha$ is between $0.01$ and $0.1$. We show the spectral gap of $M(t_n)$ as a function of time for $\alpha=0.01$, $\alpha=0.05$, and $\alpha=0.1$ in Fig.~\ref{fig:Workplace_hospital}(a). In this figure, we also show the number of events at each time. As in the \textsc{Reality Mining} data set, we observe that the change of the spectral gap is not related directly to the numbers of events. Additionally, we see that the spectral gap of $M(t_n)$ increases suddenly {from that of $M(t_{n-1})$} at a small number of times {$t_n$}; the spectral gap of $M(t_n)$ depends non-monotonically on $\alpha$ at these times. In this data set, at $t_n=885{,}520$, the spectral gap of $M(t_n)$ is largest for $\alpha=0.01$, second largest for $\alpha=0.05$, and smallest for $\alpha=0.1$. However, the spectral gaps of $M(t_n)$ for $\alpha=0.05$ and $\alpha=0.1$ exceed that for $\alpha=0.01$ at $t_n=892{,}280$, resulting in the non-monotonic behavior with respect to the decay rate $\alpha$ that we observed in Fig.~\ref{fig:comparison_six_data_sets}. We show the time series of ${\|\bm w_2\|}/{\|\bm v_2\|}$ in Fig.~\ref{fig:Workplace_hospital}(b). The decrease in ${\|\bm w_2\|}/{\|\bm v_2\|}$ at $t_n=885{,}520$ is largest for $\alpha=0.01$, second largest for $\alpha=0.05$, and smallest for $\alpha=0.1$. By contrast, the decrease in ${\|\bm w_2\|}/{\|\bm v_2\|}$ at $t_n=892{,}280$ is largest for $\alpha=0.1$, second largest for $\alpha=0.05$, and smallest for $\alpha=0.01$. Therefore, the behavior of ${\|\bm w_2\|}/{\|\bm v_2\|}$ suggests that the spectral gap of $M(T)$ depends non-monotonically on $\alpha$ in Fig.~\ref{fig:Workplace_hospital}(a).

In Fig.~\ref{fig:comparison_six_data_sets}, we also saw that the spectral gap of $M(T)$ for the \textsc{Hospital} network depends non-monotonically on $\alpha$ when $\alpha$ is between $0.001$ and $0.005$. In Fig.~\ref{fig:Workplace_hospital}(c), we show the spectral gap of $M(t_n)$ as a function of time for $\alpha=0.001$, $\alpha=0.002$, and $\alpha=0.003$. We also show the number of events at each time. 
In this data set, the spectral gap of $M(t_n)$ for $\alpha=0.003$ exceeds that for $\alpha=0.002$ before all nodes are in one network component (this occurs at $t_n=330{,}840$), resulting in the observed non-monotonicity with respect to the decay rate $\alpha$. We show the time series of ${\|\bm w_2\|}/{\|\bm v_2\|}$ in Fig.~\ref{fig:Workplace_hospital}(d). The behavior of the spectral gap of $M(t_n)$ and the behavior of ${\|\bm w_2\|}/{\|\bm v_2\|}$ are consistent with each other. For example, the increase in the spectral gap of $M(t_n)$ at $t_n=331{,}660$ is largest for $\alpha=0.001$, second largest for $\alpha=0.002$, and smallest for $\alpha=0.003$. The decrease in ${\|\bm w_2\|}/{\|\bm v_2\|}$ at $t_n=331{,}660$ is largest for $\alpha=0.001$, second largest for $\alpha=0.002$, and smallest for $\alpha=0.003$.


\bibliography{references5}

\providecommand{\noopsort}[1]{}\providecommand{\singleletter}[1]{#1}%
\begin{thebibliography}{56}%
\makeatletter
\providecommand \@ifxundefined [1]{%
 \@ifx{#1\undefined}
}%
\providecommand \@ifnum [1]{%
 \ifnum #1\expandafter \@firstoftwo
 \else \expandafter \@secondoftwo
 \fi
}%
\providecommand \@ifx [1]{%
 \ifx #1\expandafter \@firstoftwo
 \else \expandafter \@secondoftwo
 \fi
}%
\providecommand \natexlab [1]{#1}%
\providecommand \enquote  [1]{``#1''}%
\providecommand \bibnamefont  [1]{#1}%
\providecommand \bibfnamefont [1]{#1}%
\providecommand \citenamefont [1]{#1}%
\providecommand \href@noop [0]{\@secondoftwo}%
\providecommand \href [0]{\begingroup \@sanitize@url \@href}%
\providecommand \@href[1]{\@@startlink{#1}\@@href}%
\providecommand \@@href[1]{\endgroup#1\@@endlink}%
\providecommand \@sanitize@url [0]{\catcode `\\12\catcode `\$12\catcode
  `\&12\catcode `\#12\catcode `\^12\catcode `\_12\catcode `\%12\relax}%
\providecommand \@@startlink[1]{}%
\providecommand \@@endlink[0]{}%
\providecommand \url  [0]{\begingroup\@sanitize@url \@url }%
\providecommand \@url [1]{\endgroup\@href {#1}{\urlprefix }}%
\providecommand \urlprefix  [0]{URL }%
\providecommand \Eprint [0]{\href }%
\providecommand \doibase [0]{https://doi.org/}%
\providecommand \selectlanguage [0]{\@gobble}%
\providecommand \bibinfo  [0]{\@secondoftwo}%
\providecommand \bibfield  [0]{\@secondoftwo}%
\providecommand \translation [1]{[#1]}%
\providecommand \BibitemOpen [0]{}%
\providecommand \bibitemStop [0]{}%
\providecommand \bibitemNoStop [0]{.\EOS\space}%
\providecommand \EOS [0]{\spacefactor3000\relax}%
\providecommand \BibitemShut  [1]{\csname bibitem#1\endcsname}%
\let\auto@bib@innerbib\@empty
\bibitem [{\citenamefont {Newman}(2018)}]{newman2018networks}%
  \BibitemOpen
  \bibfield  {author} {\bibinfo {author} {\bibfnamefont {M.~E.~J.}\
  \bibnamefont {Newman}},\ }\href@noop {} {\emph {\bibinfo {title}
  {Networks}}}\ (\bibinfo  {publisher} {Oxford University Press, Oxford, UK,
  Second edition},\ \bibinfo {year} {2018})\BibitemShut {NoStop}%
\bibitem [{\citenamefont {Barrat}\ \emph {et~al.}(2008)\citenamefont {Barrat},
  \citenamefont {Barthelemy},\ and\ \citenamefont
  {Vespignani}}]{barrat2008dynamical}%
  \BibitemOpen
  \bibfield  {author} {\bibinfo {author} {\bibfnamefont {A.}~\bibnamefont
  {Barrat}}, \bibinfo {author} {\bibfnamefont {M.}~\bibnamefont {Barthelemy}},\
  and\ \bibinfo {author} {\bibfnamefont {A.}~\bibnamefont {Vespignani}},\
  }\href@noop {} {\emph {\bibinfo {title} {Dynamical Processes on Complex
  Networks}}}\ (\bibinfo  {publisher} {Cambridge University Press, Cambridge,
  UK},\ \bibinfo {year} {2008})\BibitemShut {NoStop}%
\bibitem [{\citenamefont {Boccaletti}\ \emph {et~al.}(2006)\citenamefont
  {Boccaletti}, \citenamefont {Latora}, \citenamefont {Moreno}, \citenamefont
  {Chavez},\ and\ \citenamefont {Hwang}}]{boccaletti2006complex}%
  \BibitemOpen
  \bibfield  {author} {\bibinfo {author} {\bibfnamefont {S.}~\bibnamefont
  {Boccaletti}}, \bibinfo {author} {\bibfnamefont {V.}~\bibnamefont {Latora}},
  \bibinfo {author} {\bibfnamefont {Y.}~\bibnamefont {Moreno}}, \bibinfo
  {author} {\bibfnamefont {M.}~\bibnamefont {Chavez}},\ and\ \bibinfo {author}
  {\bibfnamefont {D.-U.}\ \bibnamefont {Hwang}},\ }\bibfield  {title} {\bibinfo
  {title} {Complex networks: Structure and dynamics},\ }\href@noop {}
  {\bibfield  {journal} {\bibinfo  {journal} {Phys. Rep.}\ }\textbf {\bibinfo
  {volume} {424}},\ \bibinfo {pages} {175} (\bibinfo {year}
  {2006})}\BibitemShut {NoStop}%
\bibitem [{\citenamefont {Porter}\ and\ \citenamefont
  {Gleeson}(2016)}]{porter2016dynamical}%
  \BibitemOpen
  \bibfield  {author} {\bibinfo {author} {\bibfnamefont {M.~A.}\ \bibnamefont
  {Porter}}\ and\ \bibinfo {author} {\bibfnamefont {J.~P.}\ \bibnamefont
  {Gleeson}},\ }\bibfield  {title} {\bibinfo {title} {\emph{Dynamical Systems
  on Networks: A Tutorial}},\ }\href@noop {} {\bibfield  {journal} {\bibinfo
  {journal} {Frontiers in Applied Dynamical Systems: Reviews and Tutorials
  (Springer International Publishing, Cham, Switzerland), Vol. 4}\ } (\bibinfo
  {year} {2016})}\BibitemShut {NoStop}%
\bibitem [{\citenamefont {Noorazar}\ \emph {et~al.}(2020)\citenamefont
  {Noorazar}, \citenamefont {Vixie}, \citenamefont {Talebanpour},\ and\
  \citenamefont {Hu}}]{noor2020}%
  \BibitemOpen
  \bibfield  {author} {\bibinfo {author} {\bibfnamefont {H.}~\bibnamefont
  {Noorazar}}, \bibinfo {author} {\bibfnamefont {K.~R.}\ \bibnamefont {Vixie}},
  \bibinfo {author} {\bibfnamefont {A.}~\bibnamefont {Talebanpour}},\ and\
  \bibinfo {author} {\bibfnamefont {Y.}~\bibnamefont {Hu}},\ }\bibfield
  {title} {\bibinfo {title} {From classical to modern opinion dynamics},\
  }\href@noop {} {\bibfield  {journal} {\bibinfo  {journal} {Int. J. Mod. Phys.
  C}\ }\textbf {\bibinfo {volume} {31}},\ \bibinfo {pages} {2050101} (\bibinfo
  {year} {2020})}\BibitemShut {NoStop}%
\bibitem [{\citenamefont {Castellano}\ \emph {et~al.}(2009)\citenamefont
  {Castellano}, \citenamefont {Fortunato},\ and\ \citenamefont
  {Loreto}}]{castellano2009statistical}%
  \BibitemOpen
  \bibfield  {author} {\bibinfo {author} {\bibfnamefont {C.}~\bibnamefont
  {Castellano}}, \bibinfo {author} {\bibfnamefont {S.}~\bibnamefont
  {Fortunato}},\ and\ \bibinfo {author} {\bibfnamefont {V.}~\bibnamefont
  {Loreto}},\ }\bibfield  {title} {\bibinfo {title} {Statistical physics of
  social dynamics},\ }\href@noop {} {\bibfield  {journal} {\bibinfo  {journal}
  {Rev. Mod. Phys.}\ }\textbf {\bibinfo {volume} {81}},\ \bibinfo {pages} {591}
  (\bibinfo {year} {2009})}\BibitemShut {NoStop}%
\bibitem [{\citenamefont {Baronchelli}(2018)}]{baronchelli2018emergence}%
  \BibitemOpen
  \bibfield  {author} {\bibinfo {author} {\bibfnamefont {A.}~\bibnamefont
  {Baronchelli}},\ }\bibfield  {title} {\bibinfo {title} {The emergence of
  consensus: {A} primer},\ }\href@noop {} {\bibfield  {journal} {\bibinfo
  {journal} {R. Soc. Open Sci.}\ }\textbf {\bibinfo {volume} {5}},\ \bibinfo
  {pages} {172189} (\bibinfo {year} {2018})}\BibitemShut {NoStop}%
\bibitem [{\citenamefont {Sen}\ and\ \citenamefont
  {Chakrabarti}(2014)}]{sen2014sociophysics}%
  \BibitemOpen
  \bibfield  {author} {\bibinfo {author} {\bibfnamefont {P.}~\bibnamefont
  {Sen}}\ and\ \bibinfo {author} {\bibfnamefont {B.~K.}\ \bibnamefont
  {Chakrabarti}},\ }\href@noop {} {\emph {\bibinfo {title} {Sociophysics: An
  Introduction}}}\ (\bibinfo  {publisher} {Oxford University Press, Oxford,
  UK},\ \bibinfo {year} {2014})\BibitemShut {NoStop}%
\bibitem [{\citenamefont {S{\^\i}rbu}\ \emph {et~al.}(2017)\citenamefont
  {S{\^\i}rbu}, \citenamefont {Loreto}, \citenamefont {Servedio},\ and\
  \citenamefont {Tria}}]{sirbu2017opinion}%
  \BibitemOpen
  \bibfield  {author} {\bibinfo {author} {\bibfnamefont {A.}~\bibnamefont
  {S{\^\i}rbu}}, \bibinfo {author} {\bibfnamefont {V.}~\bibnamefont {Loreto}},
  \bibinfo {author} {\bibfnamefont {V.~D.}\ \bibnamefont {Servedio}},\ and\
  \bibinfo {author} {\bibfnamefont {F.}~\bibnamefont {Tria}},\ }\bibfield
  {title} {\bibinfo {title} {Opinion dynamics: models, extensions and external
  effects},\ }in\ \href@noop {} {\emph {\bibinfo {booktitle} {Participatory
  Sensing, Opinions and Collective Awareness}}}\ (\bibinfo  {publisher}
  {Springer International Publishing, Cham, Switzerland},\ \bibinfo {year}
  {2017})\ pp.\ \bibinfo {pages} {363--401}\BibitemShut {NoStop}%
\bibitem [{\citenamefont {Acemo{\u{g}}lu}\ \emph {et~al.}(2013)\citenamefont
  {Acemo{\u{g}}lu}, \citenamefont {Como}, \citenamefont {Fagnani},\ and\
  \citenamefont {Ozdaglar}}]{acemouglu2013opinion}%
  \BibitemOpen
  \bibfield  {author} {\bibinfo {author} {\bibfnamefont {D.}~\bibnamefont
  {Acemo{\u{g}}lu}}, \bibinfo {author} {\bibfnamefont {G.}~\bibnamefont
  {Como}}, \bibinfo {author} {\bibfnamefont {F.}~\bibnamefont {Fagnani}},\ and\
  \bibinfo {author} {\bibfnamefont {A.}~\bibnamefont {Ozdaglar}},\ }\bibfield
  {title} {\bibinfo {title} {Opinion fluctuations and disagreement in social
  networks},\ }\href@noop {} {\bibfield  {journal} {\bibinfo  {journal} {Math.
  Oper. Res.}\ }\textbf {\bibinfo {volume} {38}},\ \bibinfo {pages} {1}
  (\bibinfo {year} {2013})}\BibitemShut {NoStop}%
\bibitem [{\citenamefont {Brooks}\ and\ \citenamefont
  {Porter}(2020)}]{brooks2020}%
  \BibitemOpen
  \bibfield  {author} {\bibinfo {author} {\bibfnamefont {H.~Z.}\ \bibnamefont
  {Brooks}}\ and\ \bibinfo {author} {\bibfnamefont {M.~A.}\ \bibnamefont
  {Porter}},\ }\bibfield  {title} {\bibinfo {title} {A model for the influence
  of media on the ideology of content in online social networks},\ }\href
  {https://doi.org/10.1103/PhysRevResearch.2.023041} {\bibfield  {journal}
  {\bibinfo  {journal} {Phys. Rev. Research}\ }\textbf {\bibinfo {volume}
  {2}},\ \bibinfo {pages} {023041} (\bibinfo {year} {2020})}\BibitemShut
  {NoStop}%
\bibitem [{\citenamefont {Holme}\ and\ \citenamefont
  {Saram{\"a}ki}(2012)}]{holme2012temporal}%
  \BibitemOpen
  \bibfield  {author} {\bibinfo {author} {\bibfnamefont {P.}~\bibnamefont
  {Holme}}\ and\ \bibinfo {author} {\bibfnamefont {J.}~\bibnamefont
  {Saram{\"a}ki}},\ }\bibfield  {title} {\bibinfo {title} {Temporal networks},\
  }\href@noop {} {\bibfield  {journal} {\bibinfo  {journal} {Phys. Rep.}\
  }\textbf {\bibinfo {volume} {519}},\ \bibinfo {pages} {97} (\bibinfo {year}
  {2012})}\BibitemShut {NoStop}%
\bibitem [{\citenamefont {Holme}(2015)}]{holme2015modern}%
  \BibitemOpen
  \bibfield  {author} {\bibinfo {author} {\bibfnamefont {P.}~\bibnamefont
  {Holme}},\ }\bibfield  {title} {\bibinfo {title} {Modern temporal network
  theory: {A} colloquium},\ }\href@noop {} {\bibfield  {journal} {\bibinfo
  {journal} {Eur. Phys. J. B}\ }\textbf {\bibinfo {volume} {88}},\ \bibinfo
  {pages} {234} (\bibinfo {year} {2015})}\BibitemShut {NoStop}%
\bibitem [{\citenamefont {Holme}\ and\ \citenamefont
  {Saram{\"a}ki~(eds.)}(2019)}]{holme2019temporal}%
  \BibitemOpen
  \bibfield  {author} {\bibinfo {author} {\bibfnamefont {P.}~\bibnamefont
  {Holme}}\ and\ \bibinfo {author} {\bibfnamefont {J.}~\bibnamefont
  {Saram{\"a}ki~(eds.)}},\ }\href@noop {} {\emph {\bibinfo {title} {Temporal
  Network Theory}}}\ (\bibinfo  {publisher} {Springer International Publishing,
  Cham, Switzerland},\ \bibinfo {year} {2019})\BibitemShut {NoStop}%
\bibitem [{\citenamefont {Masuda}\ and\ \citenamefont
  {Lambiotte}(2020)}]{masuda2016guidance}%
  \BibitemOpen
  \bibfield  {author} {\bibinfo {author} {\bibfnamefont {N.}~\bibnamefont
  {Masuda}}\ and\ \bibinfo {author} {\bibfnamefont {R.}~\bibnamefont
  {Lambiotte}},\ }\href@noop {} {\emph {\bibinfo {title} {A Guide to Temporal
  Networks}}}\ (\bibinfo  {publisher} {World Scientific Publishing, Singapore,
  second edition},\ \bibinfo {year} {2020})\BibitemShut {NoStop}%
\bibitem [{\citenamefont {Fern{\'a}ndez-Gracia}\ \emph
  {et~al.}(2011)\citenamefont {Fern{\'a}ndez-Gracia}, \citenamefont
  {Egu{\'\i}luz},\ and\ \citenamefont {San~Miguel}}]{fernandez2011update}%
  \BibitemOpen
  \bibfield  {author} {\bibinfo {author} {\bibfnamefont {J.}~\bibnamefont
  {Fern{\'a}ndez-Gracia}}, \bibinfo {author} {\bibfnamefont {V.~M.}\
  \bibnamefont {Egu{\'\i}luz}},\ and\ \bibinfo {author} {\bibfnamefont
  {M.}~\bibnamefont {San~Miguel}},\ }\bibfield  {title} {\bibinfo {title}
  {Update rules and interevent time distributions: {S}low ordering versus no
  ordering in the voter model},\ }\href@noop {} {\bibfield  {journal} {\bibinfo
   {journal} {Phys. Rev. E}\ }\textbf {\bibinfo {volume} {84}},\ \bibinfo
  {pages} {015103} (\bibinfo {year} {2011})}\BibitemShut {NoStop}%
\bibitem [{\citenamefont {Masuda}\ \emph {et~al.}(2013)\citenamefont {Masuda},
  \citenamefont {Klemm},\ and\ \citenamefont
  {Egu{\'\i}luz}}]{masuda2013temporal}%
  \BibitemOpen
  \bibfield  {author} {\bibinfo {author} {\bibfnamefont {N.}~\bibnamefont
  {Masuda}}, \bibinfo {author} {\bibfnamefont {K.}~\bibnamefont {Klemm}},\ and\
  \bibinfo {author} {\bibfnamefont {V.~M.}\ \bibnamefont {Egu{\'\i}luz}},\
  }\bibfield  {title} {\bibinfo {title} {Temporal networks: {S}lowing down
  diffusion by long lasting interactions},\ }\href@noop {} {\bibfield
  {journal} {\bibinfo  {journal} {Phys. Rev. Lett.}\ }\textbf {\bibinfo
  {volume} {111}},\ \bibinfo {pages} {188701} (\bibinfo {year}
  {2013})}\BibitemShut {NoStop}%
\bibitem [{\citenamefont {Olfati-Saber}\ \emph {et~al.}(2007)\citenamefont
  {Olfati-Saber}, \citenamefont {Fax},\ and\ \citenamefont
  {Murray}}]{olfati2007consensus}%
  \BibitemOpen
  \bibfield  {author} {\bibinfo {author} {\bibfnamefont {R.}~\bibnamefont
  {Olfati-Saber}}, \bibinfo {author} {\bibfnamefont {J.~A.}\ \bibnamefont
  {Fax}},\ and\ \bibinfo {author} {\bibfnamefont {R.~M.}\ \bibnamefont
  {Murray}},\ }\bibfield  {title} {\bibinfo {title} {Consensus and cooperation
  in networked multi-agent systems},\ }\href@noop {} {\bibfield  {journal}
  {\bibinfo  {journal} {Proc. IEEE}\ }\textbf {\bibinfo {volume} {95}},\
  \bibinfo {pages} {215} (\bibinfo {year} {2007})}\BibitemShut {NoStop}%
\bibitem [{\citenamefont {Takaguchi}\ \emph {et~al.}(2013)\citenamefont
  {Takaguchi}, \citenamefont {Masuda},\ and\ \citenamefont
  {Holme}}]{takaguchi2013bursty}%
  \BibitemOpen
  \bibfield  {author} {\bibinfo {author} {\bibfnamefont {T.}~\bibnamefont
  {Takaguchi}}, \bibinfo {author} {\bibfnamefont {N.}~\bibnamefont {Masuda}},\
  and\ \bibinfo {author} {\bibfnamefont {P.}~\bibnamefont {Holme}},\ }\bibfield
   {title} {\bibinfo {title} {Bursty communication patterns facilitate
  spreading in a threshold-based epidemic dynamics},\ }\href@noop {} {\bibfield
   {journal} {\bibinfo  {journal} {PLOS ONE}\ }\textbf {\bibinfo {volume}
  {8}},\ \bibinfo {pages} {e68629} (\bibinfo {year} {2013})}\BibitemShut
  {NoStop}%
\bibitem [{\citenamefont {Caceres}\ \emph {et~al.}(2011)\citenamefont
  {Caceres}, \citenamefont {Berger-Wolf},\ and\ \citenamefont
  {Grossman}}]{caceres2011temporal}%
  \BibitemOpen
  \bibfield  {author} {\bibinfo {author} {\bibfnamefont {R.~S.}\ \bibnamefont
  {Caceres}}, \bibinfo {author} {\bibfnamefont {T.}~\bibnamefont
  {Berger-Wolf}},\ and\ \bibinfo {author} {\bibfnamefont {R.}~\bibnamefont
  {Grossman}},\ }\bibfield  {title} {\bibinfo {title} {Temporal scale of
  processes in dynamic networks},\ }in\ \href@noop {} {\emph {\bibinfo
  {booktitle} {Proc. 2011 IEEE 11th International Conference on Data Mining
  Workshops}}}\ (\bibinfo {organization} {Institute of Electrical and
  Electronics Engineers, New York, NY, USA},\ \bibinfo {year} {2011})\ pp.\
  \bibinfo {pages} {925--932}\BibitemShut {NoStop}%
\bibitem [{\citenamefont {Krings}\ \emph {et~al.}(2012)\citenamefont {Krings},
  \citenamefont {Karsai}, \citenamefont {Bernhardsson}, \citenamefont
  {Blondel},\ and\ \citenamefont {Saram{\"a}ki}}]{krings2012effects}%
  \BibitemOpen
  \bibfield  {author} {\bibinfo {author} {\bibfnamefont {G.}~\bibnamefont
  {Krings}}, \bibinfo {author} {\bibfnamefont {M.}~\bibnamefont {Karsai}},
  \bibinfo {author} {\bibfnamefont {S.}~\bibnamefont {Bernhardsson}}, \bibinfo
  {author} {\bibfnamefont {V.~D.}\ \bibnamefont {Blondel}},\ and\ \bibinfo
  {author} {\bibfnamefont {J.}~\bibnamefont {Saram{\"a}ki}},\ }\bibfield
  {title} {\bibinfo {title} {Effects of time window size and placement on the
  structure of an aggregated communication network},\ }\href@noop {} {\bibfield
   {journal} {\bibinfo  {journal} {EPJ Data Sci.}\ }\textbf {\bibinfo {volume}
  {1}},\ \bibinfo {pages} {4} (\bibinfo {year} {2012})}\BibitemShut {NoStop}%
\bibitem [{\citenamefont {Liljeros}\ \emph {et~al.}(2007)\citenamefont
  {Liljeros}, \citenamefont {Giesecke},\ and\ \citenamefont
  {Holme}}]{liljeros2007contact}%
  \BibitemOpen
  \bibfield  {author} {\bibinfo {author} {\bibfnamefont {F.}~\bibnamefont
  {Liljeros}}, \bibinfo {author} {\bibfnamefont {J.}~\bibnamefont {Giesecke}},\
  and\ \bibinfo {author} {\bibfnamefont {P.}~\bibnamefont {Holme}},\ }\bibfield
   {title} {\bibinfo {title} {The contact network of inpatients in a regional
  healthcare system. a longitudinal case study},\ }\href@noop {} {\bibfield
  {journal} {\bibinfo  {journal} {Math. Popul. Stud.}\ }\textbf {\bibinfo
  {volume} {14}},\ \bibinfo {pages} {269} (\bibinfo {year} {2007})}\BibitemShut
  {NoStop}%
\bibitem [{\citenamefont {Psorakis}\ \emph {et~al.}(2012)\citenamefont
  {Psorakis}, \citenamefont {Roberts}, \citenamefont {Rezek},\ and\
  \citenamefont {Sheldon}}]{psorakis2012inferring}%
  \BibitemOpen
  \bibfield  {author} {\bibinfo {author} {\bibfnamefont {I.}~\bibnamefont
  {Psorakis}}, \bibinfo {author} {\bibfnamefont {S.~J.}\ \bibnamefont
  {Roberts}}, \bibinfo {author} {\bibfnamefont {I.}~\bibnamefont {Rezek}},\
  and\ \bibinfo {author} {\bibfnamefont {B.~C.}\ \bibnamefont {Sheldon}},\
  }\bibfield  {title} {\bibinfo {title} {Inferring social network structure in
  ecological systems from spatio-temporal data streams},\ }\href@noop {}
  {\bibfield  {journal} {\bibinfo  {journal} {J. R. Soc. Interface}\ }\textbf
  {\bibinfo {volume} {9}},\ \bibinfo {pages} {3055} (\bibinfo {year}
  {2012})}\BibitemShut {NoStop}%
\bibitem [{\citenamefont {Beguerisse~D{\'\i}az}\ \emph
  {et~al.}(2010)\citenamefont {Beguerisse~D{\'\i}az}, \citenamefont {Porter},\
  and\ \citenamefont {Onnela}}]{beguerisse2010competition}%
  \BibitemOpen
  \bibfield  {author} {\bibinfo {author} {\bibfnamefont {M.}~\bibnamefont
  {Beguerisse~D{\'\i}az}}, \bibinfo {author} {\bibfnamefont {M.~A.}\
  \bibnamefont {Porter}},\ and\ \bibinfo {author} {\bibfnamefont {J.-P.}\
  \bibnamefont {Onnela}},\ }\bibfield  {title} {\bibinfo {title} {Competition
  for popularity in bipartite networks},\ }\href@noop {} {\bibfield  {journal}
  {\bibinfo  {journal} {Chaos}\ }\textbf {\bibinfo {volume} {20}},\ \bibinfo
  {pages} {043101} (\bibinfo {year} {2010})}\BibitemShut {NoStop}%
\bibitem [{\citenamefont {Karsai}\ \emph {et~al.}(2018)\citenamefont {Karsai},
  \citenamefont {Jo},\ and\ \citenamefont {Kaski}}]{karsai2018bursty}%
  \BibitemOpen
  \bibfield  {author} {\bibinfo {author} {\bibfnamefont {M.}~\bibnamefont
  {Karsai}}, \bibinfo {author} {\bibfnamefont {H.-H.}\ \bibnamefont {Jo}},\
  and\ \bibinfo {author} {\bibfnamefont {K.}~\bibnamefont {Kaski}},\
  }\href@noop {} {\emph {\bibinfo {title} {Bursty Human Dynamics}}}\ (\bibinfo
  {publisher} {Springer International Publishing, Cham, Switzerland},\ \bibinfo
  {year} {2018})\BibitemShut {NoStop}%
\bibitem [{\citenamefont {Kivel{\"a}}\ and\ \citenamefont
  {Porter}(2015)}]{kivela2015estimating}%
  \BibitemOpen
  \bibfield  {author} {\bibinfo {author} {\bibfnamefont {M.}~\bibnamefont
  {Kivel{\"a}}}\ and\ \bibinfo {author} {\bibfnamefont {M.~A.}\ \bibnamefont
  {Porter}},\ }\bibfield  {title} {\bibinfo {title} {Estimating interevent time
  distributions from finite observation periods in communication networks},\
  }\href@noop {} {\bibfield  {journal} {\bibinfo  {journal} {Phys. Rev. E}\
  }\textbf {\bibinfo {volume} {92}},\ \bibinfo {pages} {052813} (\bibinfo
  {year} {2015})}\BibitemShut {NoStop}%
\bibitem [{\citenamefont {Moinet}\ \emph {et~al.}(2015)\citenamefont {Moinet},
  \citenamefont {Starnini},\ and\ \citenamefont
  {Pastor-Satorras}}]{moinet2015burstiness}%
  \BibitemOpen
  \bibfield  {author} {\bibinfo {author} {\bibfnamefont {A.}~\bibnamefont
  {Moinet}}, \bibinfo {author} {\bibfnamefont {M.}~\bibnamefont {Starnini}},\
  and\ \bibinfo {author} {\bibfnamefont {R.}~\bibnamefont {Pastor-Satorras}},\
  }\bibfield  {title} {\bibinfo {title} {Burstiness and aging in social
  temporal networks},\ }\href@noop {} {\bibfield  {journal} {\bibinfo
  {journal} {Phys. Rev. Lett.}\ }\textbf {\bibinfo {volume} {114}},\ \bibinfo
  {pages} {108701} (\bibinfo {year} {2015})}\BibitemShut {NoStop}%
\bibitem [{\citenamefont {Ahmad}\ \emph {et~al.}(2021)\citenamefont {Ahmad},
  \citenamefont {Porter},\ and\ \citenamefont
  {Beguerisse-D{\'\i}az}}]{ahmad2018tie}%
  \BibitemOpen
  \bibfield  {author} {\bibinfo {author} {\bibfnamefont {W.}~\bibnamefont
  {Ahmad}}, \bibinfo {author} {\bibfnamefont {M.~A.}\ \bibnamefont {Porter}},\
  and\ \bibinfo {author} {\bibfnamefont {M.}~\bibnamefont
  {Beguerisse-D{\'\i}az}},\ }\bibfield  {title} {\bibinfo {title} {Tie-decay
  networks in continuous time and eigenvector-based centralities},\ }\href@noop
  {} {\bibfield  {journal} {\bibinfo  {journal} {IEEE Transactions on Network
  Science and Engineering}\ } (\bibinfo {year} {2021})},\ \bibinfo {note}
  {\url{DOI:10.1109/TNSE.2021.3071429}}\BibitemShut {NoStop}%
\bibitem [{\citenamefont {Zuo}\ and\ \citenamefont
  {Porter}(2021)}]{zuo2019models}%
  \BibitemOpen
  \bibfield  {author} {\bibinfo {author} {\bibfnamefont {X.}~\bibnamefont
  {Zuo}}\ and\ \bibinfo {author} {\bibfnamefont {M.~A.}\ \bibnamefont
  {Porter}},\ }\bibfield  {title} {\bibinfo {title} {Models of continuous-time
  networks with tie decay, diffusion, and convection},\ }\href@noop {}
  {\bibfield  {journal} {\bibinfo  {journal} {Phys. Rev. E}\ }\textbf {\bibinfo
  {volume} {103}},\ \bibinfo {pages} {022304} (\bibinfo {year}
  {2021})}\BibitemShut {NoStop}%
\bibitem [{\citenamefont {Mirzaev}\ and\ \citenamefont
  {Gunawardena}(2013)}]{mirzaev2013laplacian}%
  \BibitemOpen
  \bibfield  {author} {\bibinfo {author} {\bibfnamefont {I.}~\bibnamefont
  {Mirzaev}}\ and\ \bibinfo {author} {\bibfnamefont {J.}~\bibnamefont
  {Gunawardena}},\ }\bibfield  {title} {\bibinfo {title} {Laplacian dynamics on
  general graphs},\ }\href@noop {} {\bibfield  {journal} {\bibinfo  {journal}
  {Bull. Math. Biol.}\ }\textbf {\bibinfo {volume} {75}},\ \bibinfo {pages}
  {2118} (\bibinfo {year} {2013})}\BibitemShut {NoStop}%
\bibitem [{\citenamefont {Urena}\ \emph {et~al.}(2019)\citenamefont {Urena},
  \citenamefont {Kou}, \citenamefont {Dong}, \citenamefont {Chiclana},\ and\
  \citenamefont {Herrera-Viedma}}]{urena2019review}%
  \BibitemOpen
  \bibfield  {author} {\bibinfo {author} {\bibfnamefont {R.}~\bibnamefont
  {Urena}}, \bibinfo {author} {\bibfnamefont {G.}~\bibnamefont {Kou}}, \bibinfo
  {author} {\bibfnamefont {Y.}~\bibnamefont {Dong}}, \bibinfo {author}
  {\bibfnamefont {F.}~\bibnamefont {Chiclana}},\ and\ \bibinfo {author}
  {\bibfnamefont {E.}~\bibnamefont {Herrera-Viedma}},\ }\bibfield  {title}
  {\bibinfo {title} {A review on trust propagation and opinion dynamics in
  social networks and group decision making frameworks},\ }\href@noop {}
  {\bibfield  {journal} {\bibinfo  {journal} {Inf. Sci.}\ }\textbf {\bibinfo
  {volume} {478}},\ \bibinfo {pages} {461} (\bibinfo {year}
  {2019})}\BibitemShut {NoStop}%
\bibitem [{\citenamefont {Bullo}(2020)}]{FB-LNS}%
  \BibitemOpen
  \bibfield  {author} {\bibinfo {author} {\bibfnamefont {F.}~\bibnamefont
  {Bullo}},\ }\href@noop {} {\emph {\bibinfo {title} {Lectures on Network
  Systems}}},\ \bibinfo {edition} {{1.4}}\ ed.\ (\bibinfo  {publisher} {Kindle
  Direct Publishing},\ \bibinfo {year} {2020})\BibitemShut {NoStop}%
\bibitem [{\citenamefont {DeGroot}(1974)}]{degroot1974reaching}%
  \BibitemOpen
  \bibfield  {author} {\bibinfo {author} {\bibfnamefont {M.~H.}\ \bibnamefont
  {DeGroot}},\ }\bibfield  {title} {\bibinfo {title} {Reaching a consensus},\
  }\href@noop {} {\bibfield  {journal} {\bibinfo  {journal} {J. Am. Stat.
  Assoc.}\ }\textbf {\bibinfo {volume} {69}},\ \bibinfo {pages} {118} (\bibinfo
  {year} {1974})}\BibitemShut {NoStop}%
\bibitem [{\citenamefont {Gauvin}\ \emph {et~al.}(2018)\citenamefont {Gauvin},
  \citenamefont {G{\'e}nois}, \citenamefont {Karsai}, \citenamefont
  {Kivel{\"a}}, \citenamefont {Takaguchi}, \citenamefont {Valdano},\ and\
  \citenamefont {Vestergaard}}]{gauvin2018randomized}%
  \BibitemOpen
  \bibfield  {author} {\bibinfo {author} {\bibfnamefont {L.}~\bibnamefont
  {Gauvin}}, \bibinfo {author} {\bibfnamefont {M.}~\bibnamefont {G{\'e}nois}},
  \bibinfo {author} {\bibfnamefont {M.}~\bibnamefont {Karsai}}, \bibinfo
  {author} {\bibfnamefont {M.}~\bibnamefont {Kivel{\"a}}}, \bibinfo {author}
  {\bibfnamefont {T.}~\bibnamefont {Takaguchi}}, \bibinfo {author}
  {\bibfnamefont {E.}~\bibnamefont {Valdano}},\ and\ \bibinfo {author}
  {\bibfnamefont {C.~L.}\ \bibnamefont {Vestergaard}},\ }\bibfield  {title}
  {\bibinfo {title} {Randomized reference models for temporal networks},\
  }\href@noop {} {\bibfield  {journal} {\bibinfo  {journal} {arXiv preprint
  arXiv:1806.04032}\ } (\bibinfo {year} {2018})}\BibitemShut {NoStop}%
\bibitem [{\citenamefont {Isella}\ \emph {et~al.}(2011)\citenamefont {Isella},
  \citenamefont {Stehl{\'e}}, \citenamefont {Barrat}, \citenamefont {Cattuto},
  \citenamefont {Pinton},\ and\ \citenamefont {Van~den Broeck}}]{isella2011s}%
  \BibitemOpen
  \bibfield  {author} {\bibinfo {author} {\bibfnamefont {L.}~\bibnamefont
  {Isella}}, \bibinfo {author} {\bibfnamefont {J.}~\bibnamefont {Stehl{\'e}}},
  \bibinfo {author} {\bibfnamefont {A.}~\bibnamefont {Barrat}}, \bibinfo
  {author} {\bibfnamefont {C.}~\bibnamefont {Cattuto}}, \bibinfo {author}
  {\bibfnamefont {J.-F.}\ \bibnamefont {Pinton}},\ and\ \bibinfo {author}
  {\bibfnamefont {W.}~\bibnamefont {Van~den Broeck}},\ }\bibfield  {title}
  {\bibinfo {title} {What's in a crowd? {Analysis} of face-to-face behavioral
  networks},\ }\href@noop {} {\bibfield  {journal} {\bibinfo  {journal} {J.
  Theor. Biol.}\ }\textbf {\bibinfo {volume} {271}},\ \bibinfo {pages} {166}
  (\bibinfo {year} {2011})}\BibitemShut {NoStop}%
\bibitem [{\citenamefont {{SocioPatterns}}(2020{\natexlab{a}})}]{hypertext}%
  \BibitemOpen
  \bibfield  {author} {\bibinfo {author} {\bibnamefont {{SocioPatterns}}},\
  }\href@noop {} {\bibinfo {title} {Data set: Hypertext 2009 dynamic contact
  network}},\ \bibinfo {howpublished} {\url{http://www.sociopatterns.org/data
  sets/hypertext-2009-dynamic-contact-network/}} (\bibinfo {year} {Access Date:
  June 13, 2020}{\natexlab{a}})\BibitemShut {NoStop}%
\bibitem [{\citenamefont {G{\'e}nois}\ \emph {et~al.}(2015)\citenamefont
  {G{\'e}nois}, \citenamefont {Vestergaard}, \citenamefont {Fournet},
  \citenamefont {Panisson}, \citenamefont {Bonmarin},\ and\ \citenamefont
  {Barrat}}]{genois2015data}%
  \BibitemOpen
  \bibfield  {author} {\bibinfo {author} {\bibfnamefont {M.}~\bibnamefont
  {G{\'e}nois}}, \bibinfo {author} {\bibfnamefont {C.~L.}\ \bibnamefont
  {Vestergaard}}, \bibinfo {author} {\bibfnamefont {J.}~\bibnamefont
  {Fournet}}, \bibinfo {author} {\bibfnamefont {A.}~\bibnamefont {Panisson}},
  \bibinfo {author} {\bibfnamefont {I.}~\bibnamefont {Bonmarin}},\ and\
  \bibinfo {author} {\bibfnamefont {A.}~\bibnamefont {Barrat}},\ }\bibfield
  {title} {\bibinfo {title} {Data on face-to-face contacts in an office
  building suggest a low-cost vaccination strategy based on community
  linkers},\ }\href@noop {} {\bibfield  {journal} {\bibinfo  {journal} {Netw.
  Sci.}\ }\textbf {\bibinfo {volume} {3}},\ \bibinfo {pages} {326} (\bibinfo
  {year} {2015})}\BibitemShut {NoStop}%
\bibitem [{\citenamefont {{SocioPatterns}}(2020{\natexlab{b}})}]{workplace}%
  \BibitemOpen
  \bibfield  {author} {\bibinfo {author} {\bibnamefont {{SocioPatterns}}},\
  }\href@noop {} {\bibinfo {title} {Data set: Contacts in a workplace}},\
  \bibinfo {howpublished}
  {\url{http://www.sociopatterns.org/datasets/contacts-in-a-workplace/}}
  (\bibinfo {year} {Access Date: June 13, 2020}{\natexlab{b}})\BibitemShut
  {NoStop}%
\bibitem [{\citenamefont {{SocioPatterns}}(2020{\natexlab{c}})}]{hospital}%
  \BibitemOpen
  \bibfield  {author} {\bibinfo {author} {\bibnamefont {{SocioPatterns}}},\
  }\href@noop {} {\bibinfo {title} {Data set: Hospital ward dynamic contact
  network}},\ \bibinfo {howpublished}
  {\url{http://www.sociopatterns.org/datasets/hospital-ward-dynamic-contact-network/}}
  (\bibinfo {year} {Access Date: June 13, 2020}{\natexlab{c}})\BibitemShut
  {NoStop}%
\bibitem [{\citenamefont {Vanhems}\ \emph {et~al.}(2013)\citenamefont
  {Vanhems}, \citenamefont {Barrat}, \citenamefont {Cattuto}, \citenamefont
  {Pinton}, \citenamefont {Khanafer}, \citenamefont {R{\'e}gis}, \citenamefont
  {Kim}, \citenamefont {Comte},\ and\ \citenamefont
  {Voirin}}]{vanhems2013estimating}%
  \BibitemOpen
  \bibfield  {author} {\bibinfo {author} {\bibfnamefont {P.}~\bibnamefont
  {Vanhems}}, \bibinfo {author} {\bibfnamefont {A.}~\bibnamefont {Barrat}},
  \bibinfo {author} {\bibfnamefont {C.}~\bibnamefont {Cattuto}}, \bibinfo
  {author} {\bibfnamefont {J.-F.}\ \bibnamefont {Pinton}}, \bibinfo {author}
  {\bibfnamefont {N.}~\bibnamefont {Khanafer}}, \bibinfo {author}
  {\bibfnamefont {C.}~\bibnamefont {R{\'e}gis}}, \bibinfo {author}
  {\bibfnamefont {B.-A.}\ \bibnamefont {Kim}}, \bibinfo {author} {\bibfnamefont
  {B.}~\bibnamefont {Comte}},\ and\ \bibinfo {author} {\bibfnamefont
  {N.}~\bibnamefont {Voirin}},\ }\bibfield  {title} {\bibinfo {title}
  {Estimating potential infection transmission routes in hospital wards using
  wearable proximity sensors},\ }\href@noop {} {\bibfield  {journal} {\bibinfo
  {journal} {PLOS ONE}\ }\textbf {\bibinfo {volume} {8}},\ \bibinfo {pages}
  {e73970} (\bibinfo {year} {2013})}\BibitemShut {NoStop}%
\bibitem [{\citenamefont {Gemmetto}\ \emph {et~al.}(2014)\citenamefont
  {Gemmetto}, \citenamefont {Barrat},\ and\ \citenamefont
  {Cattuto}}]{gemmetto2014mitigation}%
  \BibitemOpen
  \bibfield  {author} {\bibinfo {author} {\bibfnamefont {V.}~\bibnamefont
  {Gemmetto}}, \bibinfo {author} {\bibfnamefont {A.}~\bibnamefont {Barrat}},\
  and\ \bibinfo {author} {\bibfnamefont {C.}~\bibnamefont {Cattuto}},\
  }\bibfield  {title} {\bibinfo {title} {Mitigation of infectious disease at
  school: {T}argeted class closure vs school closure},\ }\href@noop {}
  {\bibfield  {journal} {\bibinfo  {journal} {BMC Infect. Dis.}\ }\textbf
  {\bibinfo {volume} {14}},\ \bibinfo {pages} {695} (\bibinfo {year}
  {2014})}\BibitemShut {NoStop}%
\bibitem [{\citenamefont {{SocioPatterns}}(2020{\natexlab{d}})}]{primary}%
  \BibitemOpen
  \bibfield  {author} {\bibinfo {author} {\bibnamefont {{SocioPatterns}}},\
  }\href@noop {} {\bibinfo {title} {Data set: Primary school temporal network
  data}},\ \bibinfo {howpublished}
  {\url{http://www.sociopatterns.org/datasets/primary-school-temporal-network-data/}}
  (\bibinfo {year} {Access Date: June 13, 2020}{\natexlab{d}})\BibitemShut
  {NoStop}%
\bibitem [{\citenamefont {Stehl{\'e}}\ \emph {et~al.}(2011)\citenamefont
  {Stehl{\'e}}, \citenamefont {Voirin}, \citenamefont {Barrat}, \citenamefont
  {Cattuto}, \citenamefont {Isella}, \citenamefont {Pinton}, \citenamefont
  {Quaggiotto}, \citenamefont {Van~den Broeck}, \citenamefont {R{\'e}gis},
  \citenamefont {Lina} \emph {et~al.}}]{stehle2011high}%
  \BibitemOpen
  \bibfield  {author} {\bibinfo {author} {\bibfnamefont {J.}~\bibnamefont
  {Stehl{\'e}}}, \bibinfo {author} {\bibfnamefont {N.}~\bibnamefont {Voirin}},
  \bibinfo {author} {\bibfnamefont {A.}~\bibnamefont {Barrat}}, \bibinfo
  {author} {\bibfnamefont {C.}~\bibnamefont {Cattuto}}, \bibinfo {author}
  {\bibfnamefont {L.}~\bibnamefont {Isella}}, \bibinfo {author} {\bibfnamefont
  {J.-F.}\ \bibnamefont {Pinton}}, \bibinfo {author} {\bibfnamefont
  {M.}~\bibnamefont {Quaggiotto}}, \bibinfo {author} {\bibfnamefont
  {W.}~\bibnamefont {Van~den Broeck}}, \bibinfo {author} {\bibfnamefont
  {C.}~\bibnamefont {R{\'e}gis}}, \bibinfo {author} {\bibfnamefont
  {B.}~\bibnamefont {Lina}}, \emph {et~al.},\ }\bibfield  {title} {\bibinfo
  {title} {High-resolution measurements of face-to-face contact patterns in a
  primary school},\ }\href@noop {} {\bibfield  {journal} {\bibinfo  {journal}
  {PLOS ONE}\ }\textbf {\bibinfo {volume} {6}},\ \bibinfo {pages} {e23176}
  (\bibinfo {year} {2011})}\BibitemShut {NoStop}%
\bibitem [{\citenamefont {Fournet}\ and\ \citenamefont
  {Barrat}(2014)}]{fournet2014contact}%
  \BibitemOpen
  \bibfield  {author} {\bibinfo {author} {\bibfnamefont {J.}~\bibnamefont
  {Fournet}}\ and\ \bibinfo {author} {\bibfnamefont {A.}~\bibnamefont
  {Barrat}},\ }\bibfield  {title} {\bibinfo {title} {Contact patterns among
  high school students},\ }\href@noop {} {\bibfield  {journal} {\bibinfo
  {journal} {PLOS ONE}\ }\textbf {\bibinfo {volume} {9}},\ \bibinfo {pages}
  {e107878} (\bibinfo {year} {2014})}\BibitemShut {NoStop}%
\bibitem [{\citenamefont {{SocioPatterns}}(2020{\natexlab{e}})}]{highschool}%
  \BibitemOpen
  \bibfield  {author} {\bibinfo {author} {\bibnamefont {{SocioPatterns}}},\
  }\href@noop {} {\bibinfo {title} {Data set: High school dynamic contact
  networks}},\ \bibinfo {howpublished}
  {\url{http://www.sociopatterns.org/datasets/high-school-dynamic-contact-networks/}}
  (\bibinfo {year} {Access Date: June 13, 2020}{\natexlab{e}})\BibitemShut
  {NoStop}%
\bibitem [{\citenamefont {Eagle}\ and\ \citenamefont
  {Pentland}(2006)}]{eagle2006reality}%
  \BibitemOpen
  \bibfield  {author} {\bibinfo {author} {\bibfnamefont {N.}~\bibnamefont
  {Eagle}}\ and\ \bibinfo {author} {\bibfnamefont {A.~S.}\ \bibnamefont
  {Pentland}},\ }\bibfield  {title} {\bibinfo {title} {Reality mining:
  {S}ensing complex social systems},\ }\href@noop {} {\bibfield  {journal}
  {\bibinfo  {journal} {Pers. Ubiquit. Comput.}\ }\textbf {\bibinfo {volume}
  {10}},\ \bibinfo {pages} {255} (\bibinfo {year} {2006})}\BibitemShut
  {NoStop}%
\bibitem [{\citenamefont {Masuda}\ and\ \citenamefont
  {Holme}(2019)}]{masuda2019detecting}%
  \BibitemOpen
  \bibfield  {author} {\bibinfo {author} {\bibfnamefont {N.}~\bibnamefont
  {Masuda}}\ and\ \bibinfo {author} {\bibfnamefont {P.}~\bibnamefont {Holme}},\
  }\bibfield  {title} {\bibinfo {title} {Detecting sequences of system states
  in temporal networks},\ }\href@noop {} {\bibfield  {journal} {\bibinfo
  {journal} {Sci. Rep.}\ }\textbf {\bibinfo {volume} {9}},\ \bibinfo {pages}
  {795} (\bibinfo {year} {2019})}\BibitemShut {NoStop}%
\bibitem [{\citenamefont {Scholtes}\ \emph {et~al.}(2014)\citenamefont
  {Scholtes}, \citenamefont {Wider}, \citenamefont {Pfitzner}, \citenamefont
  {Garas}, \citenamefont {Tessone},\ and\ \citenamefont
  {Schweitzer}}]{scholtes2014causality}%
  \BibitemOpen
  \bibfield  {author} {\bibinfo {author} {\bibfnamefont {I.}~\bibnamefont
  {Scholtes}}, \bibinfo {author} {\bibfnamefont {N.}~\bibnamefont {Wider}},
  \bibinfo {author} {\bibfnamefont {R.}~\bibnamefont {Pfitzner}}, \bibinfo
  {author} {\bibfnamefont {A.}~\bibnamefont {Garas}}, \bibinfo {author}
  {\bibfnamefont {C.~J.}\ \bibnamefont {Tessone}},\ and\ \bibinfo {author}
  {\bibfnamefont {F.}~\bibnamefont {Schweitzer}},\ }\bibfield  {title}
  {\bibinfo {title} {Causality-driven slow-down and speed-up of diffusion in
  non-{M}arkovian temporal networks},\ }\href@noop {} {\bibfield  {journal}
  {\bibinfo  {journal} {Nat. Comm.}\ }\textbf {\bibinfo {volume} {5}},\
  \bibinfo {pages} {5024} (\bibinfo {year} {2014})}\BibitemShut {NoStop}%
\bibitem [{\citenamefont {Van~Mieghem}(2010)}]{van2010graph}%
  \BibitemOpen
  \bibfield  {author} {\bibinfo {author} {\bibfnamefont {P.}~\bibnamefont
  {Van~Mieghem}},\ }\href@noop {} {\emph {\bibinfo {title} {Graph Spectra for
  Complex Networks}}}\ (\bibinfo  {publisher} {Cambridge University Press,
  Cambridge, UK},\ \bibinfo {year} {2010})\BibitemShut {NoStop}%
\bibitem [{\citenamefont {Iribarren}\ and\ \citenamefont
  {Moro}(2009)}]{iribarren2009impact}%
  \BibitemOpen
  \bibfield  {author} {\bibinfo {author} {\bibfnamefont {J.~L.}\ \bibnamefont
  {Iribarren}}\ and\ \bibinfo {author} {\bibfnamefont {E.}~\bibnamefont
  {Moro}},\ }\bibfield  {title} {\bibinfo {title} {Impact of human activity
  patterns on the dynamics of information diffusion},\ }\href@noop {}
  {\bibfield  {journal} {\bibinfo  {journal} {Phys. Rev. Lett.}\ }\textbf
  {\bibinfo {volume} {103}},\ \bibinfo {pages} {038702} (\bibinfo {year}
  {2009})}\BibitemShut {NoStop}%
\bibitem [{\citenamefont {Karsai}\ \emph {et~al.}(2011)\citenamefont {Karsai},
  \citenamefont {Kivel{\"a}}, \citenamefont {Pan}, \citenamefont {Kaski},
  \citenamefont {Kert{\'e}sz}, \citenamefont {Barab{\'a}si},\ and\
  \citenamefont {Saram{\"a}ki}}]{karsai2011small}%
  \BibitemOpen
  \bibfield  {author} {\bibinfo {author} {\bibfnamefont {M.}~\bibnamefont
  {Karsai}}, \bibinfo {author} {\bibfnamefont {M.}~\bibnamefont {Kivel{\"a}}},
  \bibinfo {author} {\bibfnamefont {R.~K.}\ \bibnamefont {Pan}}, \bibinfo
  {author} {\bibfnamefont {K.}~\bibnamefont {Kaski}}, \bibinfo {author}
  {\bibfnamefont {J.}~\bibnamefont {Kert{\'e}sz}}, \bibinfo {author}
  {\bibfnamefont {A.-L.}\ \bibnamefont {Barab{\'a}si}},\ and\ \bibinfo {author}
  {\bibfnamefont {J.}~\bibnamefont {Saram{\"a}ki}},\ }\bibfield  {title}
  {\bibinfo {title} {Small but slow world: How network topology and burstiness
  slow down spreading},\ }\href@noop {} {\bibfield  {journal} {\bibinfo
  {journal} {Phys. Rev. E}\ }\textbf {\bibinfo {volume} {83}},\ \bibinfo
  {pages} {025102} (\bibinfo {year} {2011})}\BibitemShut {NoStop}%
\bibitem [{\citenamefont {Vazquez}\ \emph {et~al.}(2007)\citenamefont
  {Vazquez}, \citenamefont {R{\'a}cz}, \citenamefont {Luk{\'a}cs},\ and\
  \citenamefont {Barab{\'a}si}}]{vazquez2007impact}%
  \BibitemOpen
  \bibfield  {author} {\bibinfo {author} {\bibfnamefont {A.}~\bibnamefont
  {Vazquez}}, \bibinfo {author} {\bibfnamefont {B.}~\bibnamefont {R{\'a}cz}},
  \bibinfo {author} {\bibfnamefont {A.}~\bibnamefont {Luk{\'a}cs}},\ and\
  \bibinfo {author} {\bibfnamefont {A.-L.}\ \bibnamefont {Barab{\'a}si}},\
  }\bibfield  {title} {\bibinfo {title} {Impact of non-poissonian activity
  patterns on spreading processes},\ }\href@noop {} {\bibfield  {journal}
  {\bibinfo  {journal} {Phys. Rev. Lett.}\ }\textbf {\bibinfo {volume} {98}},\
  \bibinfo {pages} {158702} (\bibinfo {year} {2007})}\BibitemShut {NoStop}%
\bibitem [{\citenamefont {Masuda}(2016)}]{masuda2016accelerating}%
  \BibitemOpen
  \bibfield  {author} {\bibinfo {author} {\bibfnamefont {N.}~\bibnamefont
  {Masuda}},\ }\bibfield  {title} {\bibinfo {title} {Accelerating coordination
  in temporal networks by engineering the link order},\ }\href@noop {}
  {\bibfield  {journal} {\bibinfo  {journal} {Sci. Rep.}\ }\textbf {\bibinfo
  {volume} {6}},\ \bibinfo {pages} {22105} (\bibinfo {year}
  {2016})}\BibitemShut {NoStop}%
\bibitem [{\citenamefont {Gelardi}\ \emph {et~al.}(2021)\citenamefont
  {Gelardi}, \citenamefont {Barrat},\ and\ \citenamefont
  {Claidiere}}]{gelardi2021from}%
  \BibitemOpen
  \bibfield  {author} {\bibinfo {author} {\bibfnamefont {V.}~\bibnamefont
  {Gelardi}}, \bibinfo {author} {\bibfnamefont {A.}~\bibnamefont {Barrat}},\
  and\ \bibinfo {author} {\bibfnamefont {N.}~\bibnamefont {Claidiere}},\
  }\bibfield  {title} {\bibinfo {title} {From temporal network data to the
  dynamics of social relationships},\ }\href@noop {} {\bibfield  {journal}
  {\bibinfo  {journal} {arXiv preprint arXiv:2103.11755}\ } (\bibinfo {year}
  {2021})}\BibitemShut {NoStop}%
\bibitem [{\citenamefont {Gross}\ and\ \citenamefont
  {Blasius}(2008)}]{gross2008adaptive}%
  \BibitemOpen
  \bibfield  {author} {\bibinfo {author} {\bibfnamefont {T.}~\bibnamefont
  {Gross}}\ and\ \bibinfo {author} {\bibfnamefont {B.}~\bibnamefont
  {Blasius}},\ }\bibfield  {title} {\bibinfo {title} {Adaptive coevolutionary
  networks: a review},\ }\href@noop {} {\bibfield  {journal} {\bibinfo
  {journal} {J. R. Soc. Interface}\ }\textbf {\bibinfo {volume} {5}},\ \bibinfo
  {pages} {259} (\bibinfo {year} {2008})}\BibitemShut {NoStop}%
\bibitem [{\citenamefont {Chen}\ and\ \citenamefont
  {Porter}(2020)}]{chen2020epidemic}%
  \BibitemOpen
  \bibfield  {author} {\bibinfo {author} {\bibfnamefont {Q.}~\bibnamefont
  {Chen}}\ and\ \bibinfo {author} {\bibfnamefont {M.~A.}\ \bibnamefont
  {Porter}},\ }\bibfield  {title} {\bibinfo {title} {Epidemic thresholds of
  infectious diseases on tie-decay networks},\ }\href@noop {} {\bibfield
  {journal} {\bibinfo  {journal} {arXiv preprint arXiv:2009.12932}\ } (\bibinfo
  {year} {2020})}\BibitemShut {NoStop}%
\end{thebibliography}%

\end{document}